\documentclass{aa}  
\makeatletter
\let\aa@makecaption\@makecaption 
\makeatother

\usepackage{graphicx}
\usepackage{txfonts}
\usepackage[compatibility=false]{caption}
\usepackage{subcaption} 
\usepackage{lipsum}

\usepackage{amssymb,amsmath,mathrsfs,amsfonts,latexsym}
\usepackage{savesym}
\savesymbol{iint}
\restoresymbol{TXF}{iint}

\usepackage{booktabs}
\usepackage{siunitx} 
\usepackage{ctable}

\usepackage{xcolor}

\usepackage{verbatim}
\usepackage[version=3]{mhchem}
\usepackage{natbib}

\usepackage{placeins}  
\usepackage{lscape} 
\usepackage{hyperref}

\begin{document}

   \authorrunning{Hosseini and Soleimanpour} 
   \titlerunning{RNN-CNN Reconstruction of 21-cm Brightness Temperature}
   \title{Stacked Hybrid RNN-CNN Reconstruction of X-ray Influence on 21-cm Brightness Temperature}


   \author{S. Mobina Hosseini\inst{1}
        \and Bahareh Soleimanpour Salmasi\inst{1}\thanks{}
        }

   \institute{Department of Physics, Shahid Beheshti University, P.O. Box 19839-69411, Tehran, Iran \\
             \email{mobinahosseini954@gmail.com}
             \email{physics.bahare@gmail.com}
             \thanks{}}

   \date{Received Mar 19, 20XX}

 
\abstract
  {
    The X-ray photons substantially affect the thermal and ionization states of the intergalactic medium (IGM) during the Epoch of Reionization (EoR), thereby significantly influencing the 21-cm line observables such as its sky-averaged (global) brightness temperature. Nevertheless, the complicated dependency of astrophysical processes on a broad spectrum of parameters, including X-ray efficiency, spectral characteristics, and gas dynamics, makes precisely simulating the effect of X-ray flux challenging. Traditional approaches, including N-body and hydrodynamical simulations, are computationally intensive and struggle to explore high-dimensional parameter spaces efficiently. We present a stacked hybrid model trained on a specific simulation intended to reconstruct the effect of X-ray flux on the global 21-cm brightness temperature during the EoR. Along with Convolutional Neural Networks (CNNs), this architecture combines two substantial forms of recurrent neural networks (RNNs), Long Short-Term Memory (LSTM) and Gated Recurrent Unit (GRU), therefore enabling fast adaptation to several X-ray flux levels. Without demanding repeated simulations, this emulator preserves temporal and spatial dependencies and generalizes to unseen parameter combinations. This matter reduces computation time by a factor of one million while preserving excellent prediction accuracy of 99.93\%, facilitating studies on high-dimensional parameter inference and sensitivity with an error margin of less than 0.35 mK. Our LSTM-GRU-CNN emulator combines recurrent and convolutional architectures to enable a robust and scalable analysis of X-ray heating effects on the global 21-cm brightness temperature during the EoR.
 }

   \keywords{Intergalactic medium -- X-ray heating  -- Epoch of Reionization -- Stacked Generalization -- Hybrid Learning
               }
 \maketitle  

\section{Introduction}
\label{introduction}

The EoR is one of the universe's most transformative eras. During the EoR, the first stars, galaxies, and quasars ignited. These light sources gradually ionized the neutral hydrogen gas that had prevailed since the beginning of the Dark Ages \citep{mondal2023prospects}. This process started about 300 million years after the Big Bang \citep{pritchard201221} and was ongoing for about a billion years, between $z \simeq 15$ and $z \simeq 6$, respectively \citep{van1945radiogolven,ewen1951observation}. Ionizing photons from these sources actively generate expanding bubbles of ionized plasma within the neutral IGM \citep{furlanetto2006cosmology}. These bubbles eventually merged and completed the reionization by $z \simeq 6$. Observational confirmation of this transition comes from spectra of distant quasars at $z \simeq 6$, which show the IGM primarily transparent and ionized \citep{furlanetto2016reionization}. One of the investigative tools to study the EoR is the 21-cm hyperfine transition of neutral hydrogen. This faint radio signal carries encoded information about the gas density, temperature, and ionization state during reionization, serving as a cosmic fossil record of this important transition period \citep{morales2010reionization}.

Detecting the 21-cm signal corresponding to the Cosmic Dawn and the EoR holds great promise for unraveling the enigma of the universe's early developmental stages. In contrast with the interferometric experiments, which focus on the power spectrum of spatial oscillation clusters, such as the Hydrogen Epoch of Reionization Array (HERA) \citep{keller2023search, hera2023first}, LOw Frequency ARray (LOFAR) \citep{van2013lofar}, Murchison Widefield Array (MWA) \citep{tingay2013murchison}, Giant Metrewave Radio Telescope (GMRT) \citep{paciga2013simulation}, global signal experiments attempt to observe the 21-cm line redshifted emission or absorption feature specifically as a broad emission or absorption mechanism against the CMB. The Experiment to Detect the Global EoR Signature (EDGES) is credited with this approach, claiming to observe a potential absorption feature at 78 MHz  \(z \sim 17\) \citep{bowman2018absorption}. However, the striking amplitude of the profile and calibration systematics remain contentious. Other attempts, such as Shaped Antenna Measurement of the Background Radio Spectrum (SARAS) \citep{patra2013saras}, have utilized their own data with new antenna designs and Bayesian approaches to reduce foreground and instrumental biases, distinct from post-processing EDGES data \citep{singh2022detection}. Post processing is also emphasized in the upcoming Radio Experiment for the Analysis of Cosmic Hydrogen (REACH) \citep{de2019reach} projects, which place focus on redundant calibration.

X-ray heating of the IGM is one of the fundamental factors shaping the timeline and morphology of reionization and the global 21-cm signal \citep{dhandha2025exploiting,saxena2023constraining,mirocha2017global}. Before and throughout reionization, X-rays from X-ray binaries and Active Galactic Nuclei (AGN) heat the IGM \citep{baek2010reionization}. They affect the overall reionization process and the growth of ionized bubbles \citep{zahn2007simulations}. The efficiency of this heating depends on the X-ray energy and the IGM density \citep{venkatesan2001heating,madau2004early}. An efficient X-ray heating transfers substantial energy to the IGM, raising its temperature above the CMB \citep{venumadhav2018heating}. and driving rapid heating during early reionization. This, in turn, accelerates the expansion and coalescence of ionized bubbles \citep{chakraborty2025reionization}. These heating effects imprint measurable signatures on the 21-cm signal \citep{fialkov2017constraining}, including its power spectrum characteristics \citep{adams2023improved,mittal2022first}, while also providing constraints on cosmic heating history \citep{vsoltinsky2025prospects,shao202321}.

In addition to X-ray heating, the Ly$\alpha$ flux has a significant impact on the 21-cm signal during the EoR by facilitating the Wouthuysen-Field effect \citep{wouthuysen1952excitation, field1958excitation}, which links the spin temperature of hydrogen to the kinetic temperature of the gas \citep{semelin2007lyman}. This coupling is required for rendering the 21-cm signal detectable and for deducing characteristics of the IGM, including the ionization state and the Ly$\alpha$ background intensity \citep{hirata2006wouthuysen}. Accurate modeling of the spatially inhomogeneous Ly$\alpha$ flux, which is influenced by the clustering of the early star-forming galaxies, the minihalos, and the radiative transfer through the IGM \citep{abel1999radiative}, necessitates computationally intensive, high-resolution cosmological simulations \citep{kravtsov1997adaptive}. The simulations are required to resolve small-scale structures (e.g., $\leq 1$ cMpc) while encompassing large cosmic volumes ($\geq 100$ cMpc) to account for variations in the Ly$\alpha$ background \citep{lomba2014critical, barkana2001beginning}. This necessity renders them prohibitively expensive, even with advancements in semi-numerical codes such as 21cmFAST \citep{mesinger2016evolution, murray202021cmfast}.

Forecasting and analyzing the global 21-cm signal from the EoR depends on cosmological models and computational approaches, ranging from machine learning to numerical Bayesian statistics \citep{van2021bayesian,shimabukuro2017analysing,hogg2018data}. Semi-numerical simulations like 21cmFAST and SimFast21 \citep{mesinger201121cmfast} produce three-dimensional representations of the primordial cosmos, including the hydrogen concentrations, the ionization fronts, and the oscillations in spin temperature \citep{santos2010fast}. As an instance, \citet{hosseini2023new} integrated 21cmFAST with the observational data to refine the constraints on the stellar-to-gas mass fraction in the minihalos $\rm f_*$ and the ionizing photon escape fraction, denoted as $\rm f_{esc}$, revealing that $-3.2 <\rm f_* < -2.5$ for halos with a mass of $10^{7} \rm M_{\odot}$ to conform to the constraints of the EoR. Comparably, \citet{greig201521cmmc} utilized Markov Chain Monte Carlo (MCMC) sampling with 21cmFAST to investigate the parameter space associated with the Cosmic Dawn and the reionization. These parameters include the ionization fraction, the gas density, the kinetic temperature, and the local Ly$\alpha$ flux \citep{baek2009simulated}.

More comprehensive codes, such as the \texttt{LICORICE} (Ly$\alpha$ and Ionization Code for Radiative Transfer in Cosmological Simulations), provide higher fidelity modeling at greater computational cost. These simulations consistently represent the Ly$\alpha$ coupling and X-ray heating through the radiative transfer, essential for addressing the variations in the global 21-cm signal \citep{semelin2007lyman}. While these simulations facilitate the investigation of large parameter space (e.g., gas density, ionization fraction $\rm x_e$), their computational demands, particularly when incorporating N-body dynamics or full hydrodynamical treatments, remain a significant challenge despite recent algorithmic improvements \citep{venkatesan2001heating}.

The RNNs are optimized for time-series analysis due to their capacity to identify patterns in data sequences \citep{grossberg2013recurrent}. Thus, it can be advantageous for analyzing data from the EoR, which is inherently time-series \citep{faaique2024overview}. In contrast, CNNs excel at reconstructing spatial properties while keeping computational cost low, which offer complementary advantages for the 21-cm signal analysis \citep{purwono2022understanding, cong2014minimizing}. The collaboration between the simulations and the neural networks enables resolving parameter degeneracies (e.g., differentiating X-ray heating from Ly$\alpha$ coupling) \citep{kacprzak2022deeplss} and enhancing predictions for telescopes such as HERA \citep{neben2016hydrogen} and the prospective Square Kilometre Array (SKA) \citep{dewdney2009square}.

Recent advances in machine learning and statistical emulation techniques have significantly enhanced our ability to model and interpret the 21-cm signal from the EoR. Emulators such as \citep{breitman202421cmemu} enable rapid computation of summary observables from semi-numerical simulations. Meanwhile, novel approaches like Gaussian process regression \citep{mertens2024retrieving} and latent-space neural network encoding \citep{patil2025efficient} improve signal extraction from noisy data, particularly for the 21-cm forest. While neural networks demonstrated high potential to analyze the 21-cm signal \citep{shimabukuro2017analysing}, our work extends these analyses using a stacked hybrid approach for efficient parameter space exploration.

We present a computationally efficient hybrid model to analyze the impact of X-ray flux on the global 21-cm brightness temperature during EoR. By integrating CNNs with RNNs, the model reconstructs X-ray heating effects without requiring repeated simulations. This approach addresses the challenges of simulating the complex dependencies of astrophysical processes on various parameters, preserving spatial and temporal dynamics while enabling fast adaptation to diverse X-ray flux levels. The emulator reduces runtime by a factor of one million while maintaining 99.93\% prediction accuracy, and facilitates sensitivity studies and high-dimensional parameter inference during the EoR. The framework's adaptability makes it an excellent tool for assessing incoming SKA observations; future advancements to the machine learning components could enable applications across different cosmological concerns beyond the 21-cm signal prediction.

The fundamental equations of the 21-cm cosmology, the required background of the EoR, and the astrophysical parameters introduced in the 21SSD simulation \citep{semelin201721ssd}, which is utilized in this study, are thoroughly reviewed in \S \ref{sec:2}. In §\ref{sec:3}, we outline the foundations of the machine learning algorithms used in this study. Our approach for building the target architecture, which includes preprocessing, feature engineering, Bayesian hyperparameter tuning, and model evaluation, is described in §\ref{sec:4}. A detailed comparison of the outputs produced by these techniques with the simulation outcomes is part of the analysis. We explore the advantages of using neural networks to predict basic astrophysical and cosmological parameters of the IGM in \S \ref{sec:5}. In \S \ref{sec:6}, we present our emulator's outstanding performance in reconstructing the global 21-cm signal. 

\section{Background} 
\label{sec:2}
This section begins by exploring the fundamental definitions and equations required to understand the physics of the global 21-cm signal. We emphasize the significance of investigating the EoR and then present the parameters influencing our X-ray background.

\subsection{21-cm Theory}
\label{sec:2.1}

    The 21-cm signal emerges from the intriguing interaction between the proton's magnetic moment and the ground-state electron. Based on \citet{rybicki2024radiative}, the radiative transfer equation describes radiation propagation in the IGM. The radiative transfer equation for an infinitesimal distance $d\rm s$ is written as
    \begin{equation}
        \frac{d \rm  I_\nu}{d \rm s}=\rm -\alpha_\nu I_\nu+j_\nu,
    \label{eq1}
    \end{equation}
    \noindent where the frequency is represented by the subscript $\nu$. $\alpha_\nu$ is the absorption coefficient and $\rm I_\nu$ is the specific intensity of the incident light. A decrease in intensity of $\rm \alpha_\nu I_\nu$ occurs as the incident light travels through the IGM due to absorption. The emission coefficient per unit volume per unit solid angle is denoted by $\rm j_\nu$. The infinitesimal optical depth $d\tau_\nu$ is defined as $d\tau_\nu \equiv \alpha_\nu d\rm s$, and the source function $\rm S_\nu$ is expressed as $\rm S_\nu \equiv \frac{j_\nu}{\alpha_\nu}$ \citep{shimabukuro2023exploring}. In the thermal equilibrium, the source function $\rm S_\nu$ is expressed by the Planck function $\rm B_\nu$. Accordingly, Eq. (\ref{eq1}) can be rewritten as
    \begin{equation}
     \frac{d \rm I_\nu}{d \rm \tau_\nu}=\rm -I_\nu+B_\nu.
    \label{eq2}
    \end{equation}
    The solution of Eq. (\ref{eq2}) is as follows
    \begin{equation}
    \rm I_\nu=I_\nu(0) \exp \left(-\tau_\nu\right)+B_\nu\left[1-\exp \left(-\tau_\nu\right)\right],
    \label{eq3}
    \end{equation}
     where $\tau_\nu$ is an optical depth at frequency \(\nu\), the first component of this equation represents the extinction of the incident radiation absorbed by the IGM, while the second term reflects the extinction of the emission from the source. 

    The optical depth of the IGM for the 21-cm signal depends on several key physical parameters, as described by
\begin{equation}
     \tau_{21}=\rm \frac{3c^{3}\hbar A_{10}x_{HI}n_{H}}{16k_{B}T_{s}\nu_{0}^{2}}\frac{1}{H(z)+(1+z)\partial_{r}v_{r}},
\end{equation}
   
    \noindent here, \(\tau_{21}\) is the optical depth specific to the 21-cm transition, distinct from the general \(\tau_\nu\) in Eqs. \ref{eq1}-\ref{eq3}. $\hbar$ is the reduced Planck constant, $\rm A_{10} = 2.85\times 10^{\text{-15}} s^{\text{-1}}$ is the Einstein spontaneous emission rate coefficient \citep{field1958excitation}, $\rm x_{HI}$ denotes the neutral fraction of hydrogen, $\rm n_{H}$ is the hydrogen comoving number density, $\nu_{0} = 1420.4$ MHz is the rest-frame frequency of the 21-cm signal, $\rm H(z)$ is the Hubble parameter, and $\partial_{\rm r}\rm v_{r}$ is the comoving gradient of the peculiar velocity along the line of sight \citep{field1959attempt}.

    The Rayleigh-Jeans approximation can be employed in the low-frequency regime applicable to the 21-cm line and the thermal equilibrium state. Consequently, $\rm I_\nu= 2 k_{\mathrm{B}} T_{b} \nu^2 / c^2, I_\nu(0)=\rm 2 k_{\mathrm{B}} T_R(\nu) \nu^2 / c^2, B_\nu=\rm 2 k_{\mathrm{B}} T_{\text {ex }} \nu^2 / c^2$, correspondingly. $\rm T_b$, $\rm T_{e x}$, and $\rm T_R$ are designated as the brightness temperature, the excitation temperature, and the brightness of the radio background source, respectively. The brightness temperature is frequently utilized to quantify the apparent temperature of an object according to its measured specific intensity. Utilizing these temperatures, and for the 21-cm line, Eq. \ref{eq3} simplifies to
    \begin{equation}
    \rm T_{b}=T_R({21}) e^ {(-\tau_{21})}+T_{\mathrm{ex}}\left[1-e^ {\left(-\tau_{21}\right)}\right],
    \label{eq5}
    \end{equation}
    \noindent where the neutral hydrogen distribution primarily affects the 21-cm optical depth $\tau_{21}$, while it is also dictated by the spin temperature and peculiar velocities. The relative contributions of the $\rm T_R$ and  $\rm T_{\rm ex}$ temperatures are weighted by this optical depth. The first term represents the attenuated background radio radiation, primarily from the CMB, as it passes through the neutral hydrogen medium. The second term reflects the impact of the CMB radiation on the global 21-cm brightness temperature. In cosmology, the brightness temperature of the background is represented by the CMB temperature $\rm T_\gamma$. The CMB resembles a blackbody spectrum with an approximate temperature of $\rm 2.73(1 + z) K$. In hyperfine structure, the excitation temperature $\rm T_{\text{ex}}$ is substituted with the spin temperature $\rm T_{\mathrm{s}}$. The spin temperature $\rm T_{\mathrm{s}}$ is defined using the ratio of the number densities of hydrogen atoms in the singlet hyperfine level ($\rm n_0$) and the triplet hyperfine level ($\rm n_1$) \citep{furlanetto2006cosmology}
    \begin{equation}
    \rm \frac{n_1}{n_0}=\frac{g_1}{g_0} \exp \left(\frac{-hc}{k_{\mathrm{B}} \lambda_{21} T_{\mathrm{S}}}\right),
    \label{eq6}
    \end{equation}

   \noindent where $\rm T_{*} \equiv \frac{hc}{k_B \lambda_{21}} = 0.068 \, \text{K}$, $\rm h$ is the dimensionless Hubble constant, $\rm c$ represents the speed of light, $\rm k_B \simeq 1.38 \times 10^{-23} \, \text{J} \cdot \text{K}^{-1}$ is the Boltzmann constant, and $\rm \left(\frac{g_1}{g_0}\right) = 3$ is the ratio of the statistical degeneracy factors of the two states \citep{field1958excitation}.

    The variation between hydrogen clouds and the CMB is observed. Hence, the differential 21-cm brightness temperature is evaluated as follows \citep{furlanetto2006cosmology,scott199021}
   
    \begin{equation} \label{eq7} 
    \begin{split}
    \delta \rm T_{b}=\frac{T_{s}-T_{\gamma}}{1+z}(1+e^{-\tau_{21}})\hspace{3.2cm}\\ =\rm 27x_{HI}(1-\frac{T_{\gamma}}{T_{s}})(\frac{0.15}{\Omega_{m}}\frac{1+z}{10})^{\frac{1}{2}}(\frac{\Omega_{b}h}{0.023})mK,
    \end{split}
    \end{equation}
    where $\rm \Omega_m$ and $\rm \Omega_b$ denote the fractional energy content of matter and the baryonic matter, respectively \citep{aghanim2020planck}. 

    The spin temperature can be formulated as 
    \begin{equation} \label{eq8}
    \rm T_{s}^{-1}=\frac{T_{\gamma}^{-1}+\rm x_{\alpha}\rm T_\alpha ^{-1}+\rm x_{c}\rm T_{k}^{-1}}{1+\rm x_{\alpha}+\rm x_{c}}. 
    \end{equation}

  \noindent  In this context, the gas's kinetic temperature is $\rm T_{k}$, and its Ly$\alpha$ temperature is $\rm T_\alpha $. The coupling coefficients for the collisions scattering is $\rm x_{c}$ and the Ly$\alpha$ is $\rm x_{\alpha}$.\\
    $\rm x_{\alpha}$ can be written as
    \begin{equation}
       \rm x_\alpha \equiv \frac{P_{10}T_{*}}{A_{10}T_{\gamma}},
    \end{equation}
    
    \noindent where $\rm P_{10}=(4/27)P_\alpha$ is the transition rate between the hyperfine states, and $\rm P_\alpha$ is Ly$\alpha$ scattering rate \citep{field1959spin}. \\$\rm x_{c}$ can be calculated as follows 
    \begin{equation} \label{eq9} 
    \begin{split}
    \rm x_{c}\equiv \frac{C_{10}T_{*}}{A_{10}T_{\gamma}}=\rm x_{c}^{HH}+\rm x_{c}^{eH}+\rm x_{c}^{pH}\hspace{4cm}\\ =\frac{\rm T_{*}}{\rm A_{10}T_{\gamma}}[\rm k_{10}^{HH}(\rm T_{k})n_{H}+k_{10}^{eH}(\rm T_{k})n_{e}+k_{10}^{pH}(\rm T_{k})n_{p}], 
    \end{split}
    \end{equation} 
    \noindent where $\rm C_{10}$ is collisional de-excitation rate of the hyperfine transition of neutral hydrogen, $\rm k_{10}^{HH}$, $\rm k_{10}^{eH}$, and $\rm k_{10}^{pH}$ reflect the rate coefficients for the three collision types \citep{zygelman2005hyperfine}. $\rm n_p$ and $\rm n_e$ show the number densities of protons and electrons in the IGM. If collisional coupling and Ly$\alpha$ scattering are efficient, then the spin temperature follows either $\rm T_{k}$ or $\rm T_\alpha $, and it is not reliant on the CMB. During the initial Ly$\alpha$ coupling era, fluctuations in the Ly$\alpha$ flux induce corresponding fluctuations in the 21-cm brightness temperature. This signal remains observable until the completion of Ly$\alpha$ coupling (i.e., when $\rm x_{\mathrm{tot}} \gg 1$), where $\rm x_{\mathrm{tot}} = x_\alpha + x_c$ is the total coupling coefficient \citep{barkana2005method}. Throughout reionization, \(\rm x_\alpha\) typically exceeds \(\rm x_c\) by orders of magnitude in the diffuse IGM (\(\rm x_\alpha \gg x_c\)), as Ly$\alpha$ photons propagate far from sources while collisions are confined to dense regions such as mini halos, but its impact is subdominant globally during reionization. For instance, at \(z=10\), \(\rm x_\alpha \sim 10^2\) in ionized bubbles, while \(\rm x_c \sim 1\) even in dense gas \citep{furlanetto2006cosmology,pritchard2007descending}.
\subsection{The Epoch of Reionization}
\label{sec:2.2}

    The EoR is a critical stage in the universe's past that includes transitioning from the dark age to the era of a bright universe \citep{zaroubi2012epoch}. This epoch started almost 300 million years after the Big Bang and lasted about 1 billion years \citep{barkana2001beginning}. Early stars and galaxies mainly emitted ultraviolet radiation, thus ionizing the surrounding hydrogen gas \citep{yoshida2003early}.

    The reionization process involved the formation of ionized bubbles around the initial light sources \citep{furlanetto2006cosmology,finkelstein2015evolution}. At first, these bubbles were relatively rare and separate; however, the further the stars and galaxies emerged, the larger the bubbles became and started to merge \citep{tilvi2020onset}. The merging of these ionized bubbles provides additional radiation sources into their vicinity, which accelerates the expansion of the ionization fronts that define their boundaries \citep{chen2023patchy}.

    The percolation process drove the formation of large ionized regions that ultimately led to a fully ionized universe \citep{furlanetto2016reionization}. This expansion initiates a chain reaction modulated by Lyman Limit Systems (LLSs), high-density regions that regulate the mean free path of Lyman continuum (LyC) photons \citep{shukla2016effects}. These LLSs maintain significant neutral hydrogen reservoirs \citep{jamieson2024thesan} and influence reionization studies by absorbing ionizing photons (neutral hydrogen column density ${\rm N}_{\text{HI}} \geq 10^{17.2}\,\text{cm}^{-2}$), scattering Ly$\alpha$ radiation, modulating IGM transparency, and reducing Ly$\alpha$ emitter visibility at $z\sim6$--$7$. Their neutral gas delays reionization completion and generates 21-cm fluctuations, making LLSs essential cosmic dawn probes. The LyC mean free path decreases significantly at $z\sim6$, with additional radiation sources including dark matter annihilation \citep{mapelli2006impact,liu2016contributions}, primordial globular clusters \citep{ricotti2002did,ma2022crash}, and micro-quasars \citep{gnedin2022modeling,madau2004early,mirabel2011stellar}. LLSs further reveal gas self-shielding properties under ionizing radiation \citep{erkal2015investigating}.

    The quasar absorption spectra and the CMB observations show that the reionization took place between $z\sim6$ and 15 \citep{mitra2015cosmic}. The Planck Collaboration \citep{aghanim2020planck} measured the Thomson scattering optical depth ($\rm \tau_e$) to be $0.054\pm0. 0092$, which would imply a most likely reionization midpoint of $z\approx7. 7\pm0. 6$ \citep{glazer2018reionization}. Some models place the end of the EoR at \(z < 6\) \citep{nasir2020observing,vsoltinsky2025prospects}. Ly\(\alpha\) forest observations further constrain late-stage reionization. While many things are becoming clearer \citep{datta2016probing}, the particulars of the EoR, its geometry, and its history remain poorly understood \citep{dayal2018early,ali201564}. The temperature anisotropies of the CMB have also been used via the Sunyaev-Zel'dovich (SZ) effect to constrain the EoR \citep{sunyaev1980microwave} with the recent explosion of large ground-based CMB observatories such as the South Pole Telescope (SPT) \citep{ruhl2004south} and the Atacama Cosmology Telescope (ACT) \citep{kosowsky2003atacama}); much of this effort has been focused on scaling up these two experiments.

    The 21-cm signal will manifest as absorption during the early phase of reionization, presumably at $z > 10$, when the total emitted X-ray energy is insufficient to elevate the IGM temperature above that of the CMB \citep{baek2010reionization}. The duration and the intensity of this absorption phase, governed by the sources' spectral energy distribution (SED), are essential. The SKA precursors capable of investigating the pertinent frequency range of 70–140 MHz could detect a significantly enhanced signal-to-noise ratio compared to subsequent periods in the EoR \citep{vazza2019detecting}. However, if the absorption phase is restricted to redshifts exceeding 15, radio frequency interference and the ionosphere will pose a challenge \citep{baek2010reionization}. The various sources of the reionization and their distinct creation histories yield significantly diverse characteristics for the 21-cm signal. As a result, forthcoming observations are required to investigate the spectrum of astrophysically probable situations by the numerical simulations \citep{eide2018epoch}.

    Three bands in the sources' SED affect the intensity of the 21-cm signal: the Lyman band, the ionizing UV band, and the soft X-ray band. The Lyman band photons are essential for decoupling the spin temperature of hydrogen from the CMB temperature via the Wouthuysen-Field effect \citep{wouthuysen1952excitation,field1958excitation}, rendering the EoR signal detectable. The ultraviolet band photons ionize the IGM, while soft X-rays can preheat the neutral gas before the ionizing front. This preheating influences whether the decoupled spin temperature is lower (weak preheating) or higher (strong preheating) than the CMB temperature \citep{baek2010reionization}. Our analysis employs the 21SSD simulation, detailed in §\ref{sec:4}. The 21-cm brightness temperature in the 21SSD simulation is mainly governed by X-ray efficiency, Lyman band emissivity, and the relative contribution of X-ray heating sources. These astrophysical parameters are presented in detail in the following subsection.

\subsection{Varying Astrophysical Parameters: Definition and Formulation}
\label{sec:2.3}
The X-rays profoundly impact the global 21-cm brightness temperature. Due to their smaller ionizing cross-section, the X-ray photons can enter the neutral hydrogen more effectively than UV photons, heating the gas above the CMB temperature \citep{santos2010fast}. The X-ray heating effect on the IGM is frequently considered homogeneous due to the long mean free path of X-rays. The X-ray flux is more intense near the sources, and this inhomogeneous flux can induce additional fluctuations in the 21-cm brightness temperature. 
The X-ray efficiency ($\rm f_X$) production during the EoR is generally described as \citep{furlanetto2006cosmology}
\begin{equation}
    \rm f_X = \frac{L_X}{3.4 \times 10^{40} \cdot \left( \frac{\text{SFR}}{1 \, M_\odot \, \text{yr}^{-1}} \right)},
\end{equation}                                           
\noindent where $\rm L_{X}$ is X-ray luminosity and the $\text{SFR}$ denotes the locally observed star formation rate.
Whenever a new source is created in the simulation, we calculate an equivalent steady-state SFR based on the mass and lifetime of the source. Then, using this information, we apply the formula to determine the X-ray luminosity, consistently using the same $\rm f_X$ value throughout the duration of the simulation.

The sources of the X-ray heating include the binaries and the AGNs. Heating from binary systems of X-ray sources produces a more uniform heating effect, which diminishes 21-cm fluctuations during the heating transition. Hard X-rays require specification in consideration of where they are used because of their high-energy and short-wavelength spectrum. In the case of soft X-rays from AGNs, on the other hand, the emission of energy and spatial distribution are usually highly localized. The soft X-rays have a spectral index of $1.6$ and an energy range of $0.1$–$2\rm keV$ \citep{fialkov2014observable,furlanetto2006cosmology}. $\rm f_{X}^{XRB}$ denotes the X-ray energy of X-ray binaries, whereas $\rm f_{X}^{AGN}$ marks the X-ray energy of active galaxies. The fraction of hard to soft X-ray emissions is given by the ratio of X-ray emissivities from X-ray binaries and the AGNs as
\begin{equation}
    \rm r_{H/S}=\frac{f_{X}^{XRB}}{f_{X}^{XRB}+f_{X}^{AGN}}.
\end{equation}
The surrounding material's obscuration and absorption of X-rays can be investigated using AGNs employing their hard to soft X-ray ratio ($\rm r_{H/S}$). Greater levels of obscuration and absorption suggested by higher hardness ratios demonstrate the presence of a dusty torus or another absorbing substance \citep{yang2015correlation}.

The local Ly$\alpha$ flux is computed by the radiative transfer calculations, excluding the effects of metal enrichment on the Lyman lines. A constant luminosity for the Lyman band is presumed, and the simulation computes the energy emitted within the specified frequency range by a stellar population of stars of varying masses, as represented in \citet{vonlanthen2011distinctive}

\begin{equation}
   \rm E(\nu_{1},\nu_{2}) = \int_{\nu_{1}}^{\nu_{2}} \int_{M_{\mathrm{min}}}^{M_{\mathrm{max}}} \zeta(M) L(M,\nu) T_{\mathrm{life}}(M) \,\it d\rm M \,\it d\nu,
\end{equation}

\noindent where $\zeta(\rm M)$ indicates the Initial Mass Function (IMF), $\rm T_{\mathrm{life}}(M)$ signifies the lifetime of a star with mass $\rm M$, and $\rm L(M,\nu)$ refers to the spectral energy distribution (energy emission per unit time and frequency). 

The Lyman band emissivity efficiency ($\rm f_{\alpha}$) is then defined as the ratio of the effectively emitted energy $\rm E_{\mathrm{eff}}$ to the theoretically available energy within the relevant frequency range \citep{semelin201721ssd}

\begin{equation}
    \rm f_{\alpha} = \frac{E_{\mathrm{eff}}(\nu_{\alpha},\nu_{\mathrm{limit}})}{E(\nu_{\alpha},\nu_{\mathrm{limit}})},
\end{equation}

 \noindent where $\nu_{\alpha}$ corresponds to the Ly$\alpha$ frequency (10.2 eV/$\rm h$) and $\nu_{\mathrm{limit}}$ represents the Lyman limit frequency (13.6 eV/$\rm h$). This efficiency factor accounts for both the intrinsic stellar properties and the intervening absorption processes in the intergalactic medium, providing a crucial link between stellar population characteristics and the resulting Ly$\alpha$ flux observable in simulations.

\section{Algorithms and Techniques}
\label{sec:3}
Data analysis has been considerably improved by recent machine learning developments. These advances consist of the invention and implementation of algorithms spotting data patterns \citep{ntampaka2019role}. Machine learning techniques are therefore beneficial for detecting trends and expanding our understanding of the physics behind specific events since they are excellent at pattern recognition in high-dimensional space \citep{dvorkin2022machine}. Furthermore, machine learning has efficiently solved data-related problems \citep{alzubi2018machine}. From only training computers to replicate human brain functions, the field has evolved into an advanced discipline extending statistics into more general areas and producing important computational theories regarding learning processes \citep{nasteski2017overview}. Although employing neural networks for data processing is not new, given major hardware developments and the development of technologies, including in-memory computing and neuromorphic circuits, it has become increasingly viable \citep{smagulova2019survey}.

\subsection{Random Forest}

Decision Trees are supervised learning models that recursively partition data into subsets based on feature values to construct a tree-like structure \citep{choi2020introduction}. This approach supports both classification and regression tasks. A decision tree is a non-parametric model whereby the feature space is divided into sections, each of which corresponds to a leaf node in the tree \citep{janitza2018computationally}. To reduce impurity measures, such as the Gini impurity for classification or the Mean Squared Error (MSE) for regression, the tree divides the data at each internal node depending on a feature $\rm x_{j}$ and a threshold $\rm t$ \citep{jaiswal2017application}. For a dataset $\mathcal{D} = \{(\rm x_{\rm i}, {\rm y}_{\rm i})\}_{\rm i=1}^{\rm N}$, the optimal split is determined by solving
\begin{equation}
    (\rm j^*, \rm t^*) = \arg\min_{\rm j, \rm t} \left( \text{Impurity}(\mathcal{D}_\text{\rm left}) + \text{Impurity}(\mathcal{D}_\text{\rm right}) \right),
\end{equation}
\noindent where $\mathcal{D}_\text{\rm left}$ and $\mathcal{D}_\text{\rm right}$ are the subsets of data resulting from the split \citep{biau2012analysis}.

Random forests (RFs) expand upon this concept by constructing numerous Decision Trees during the training process and combining their outputs to further improve the predictive accuracy and robustness \citep{resende2018survey}. They create different Decision Trees applying random subspace and bagging. Every tree learns on a bootstrapped sample of the dataset produced by randomly selecting $\rm N$ data points with replacement \citep{genuer2017random}. This helps reduce overfitting and adds variety among the trees. A random subset of features is chosen at every node of a tree to decide the optimal split. This randomness ensures that the trees are diverse and not overly similar, which improves the ensemble’s performance. For regression tasks, the final prediction $\hat{\rm y}$ for an input $\rm x$ is the average of the predictions from all $T$ trees
\begin{equation}
   \hat{\rm y} = \frac{1}{\rm T} \sum_{\rm t=1}^{\rm T} {\rm f_{\rm t}}(\rm x), 
\end{equation}
\noindent where $\rm f_{\rm t}(\rm x)$ is the prediction of the $\rm t$-th tree \citep{hasan2016feature}.

RFs offer numerous computational and practical advantages, including expedited training times \citep{amrani2018train}. The training procedure is efficient because of the simultaneous creation of trees, with predictions produced through averaging or voting, which is computationally inexpensive \citep{yaman2018comparison}. This process is termed parallel implementation, achieved by the independent construction of trees, which renders RFs scalable to large datasets \citep{shanmugasundar2021comparative}. Another benefit of utilizing RFs is the ability to predict with minimal adjustment. They depend on a limited set of hyperparameters, such as the size of the random feature subset and the number of trees \citep{cutler2012random}. Moreover, they possess an estimation of generalization error; the out-of-bag (OOB) error, calculated using data excluded from the bootstrap samples, provides an intrinsic assessment of the model's generalization efficacy \citep{breiman1996out}. RFs also provide a natural measure of feature importance, which quantifies the contribution of each feature to the model’s predictive accuracy by evaluating how much it reduces impurity across all trees \citep{speiser2019comparison}. Given these benefits, RFs represent a potent and adaptable ensemble technique that merges the straightforwardness of decision trees with the resilience of ensemble learning.

\subsection{Feedforward Neural Networks}

One of the most potent machine learning methods, neural networks provide a computationally effective means of simulating complicated, non-linear relationships in data \citep{varsamopoulos2017decoding}. They are efficient emulators in astrophysics, approximating the correlation between important astrophysical parameters like $\rm f_{\rm X}$, escape percentage of ionizing photons (${\rm f}_{\text{\rm esc}}$), and star formation efficiency ($\rm f_*$) and 21-cm observables, including power spectra \citep{schmit2018emulation}. Among the various neural network architectures, Feedforward Neural Networks (FNNs) are particularly simple and easy to use \citep{dutta2018output}. Unlike recurrent or cyclic networks, FNNs are intrinsically simple to train. They are distinguished by connections between nodes that do not form cycles, meaning that data moves in one direction, from the input layer through hidden layers to the output layer \citep{rozos2021multilayer}. A hidden layer applies non-linear transformations to the input data, enabling the network to detect and learn hierarchical features. This capability allows FNNs to model increasingly complex patterns in the data \citep{al2022multi}. Mathematically, for a given input vector $\rm x \in \mathbb{R}^n$, the output of a single neuron in a hidden layer is computed as
\begin{equation}
    {\rm z}_{\rm j} = \sigma\left(\sum_{\rm i=1}^{\rm n} \rm w_{ji} x_i + \rm b_j\right),
\end{equation}
where $\rm w_{ji}$ represents the weight connecting the $\rm i$-th input to the $\rm j$-th neuron, $\rm b_j$ is the bias term, and $\sigma(\cdot)$ is a non-linear activation function \citep{cheng2021twd}.

Training an FNN entails updating the weights and biases of its neurons to minimize a loss function $\mathcal{L}(\rm {y}, \hat{\rm {y}})$ where $\rm {y}$ is the true value and $\hat{\rm {y}}$ is the predicted output \citep{lui2019construction}. One of the most frequently employed training methods is backpropagation, which involves the propagation of defects backward through the network to update the weights \citep{lui2019construction}. During this process, the gradient of the loss function with respect to each weight is computed using the chain rule \citep{thakur2021fundamentals}
\begin{equation}
    \frac{\partial \mathcal{L}}{\partial \rm w_{ji}} = \frac{\partial \mathcal{L}}{\partial \rm z_j} \cdot \frac{\partial \rm z_j}{\partial \rm w_{ji}},
\end{equation}
and adjustments are done applying gradient descent optimization
\begin{equation}
    \rm w_{ji} \leftarrow w_{ji} - \eta \frac{\partial \mathcal{L}}{\partial w_{ji}},
\end{equation}
\noindent where $\eta$ is the learning rate \citep{xue2022ensemble}.

A deep neural network (DNN) is a more sophisticated version of an FNN that includes multiple hidden layers between the input and output layers \citep{wan2017quantum}. For a DNN with $\rm L$ hidden layers, the output of the $\rm l$-th layer ($\rm l = 1, 2, \dots, L $) can be expressed as \citep{dudek2020data}
\begin{equation}
    \rm z^{(l)} = \sigma\left(\rm W^{(l)} \rm z^{(l-1)} + b^{(l)}\right),
\end{equation}
where $\rm W^{(l)}$ is the weight matrix, $\rm b^{(l)}$ is the bias vector, and $\rm z^{(\rm l-1)}$ is the output from the previous layer. In astrophysics, neural networks are applied to tasks such as predicting galaxy properties \citep{guo2024multi} and analyzing the CMB \citep{ni2023cmb}. Their ability to handle high-dimensional data and complex relationships makes them indispensable tools for modern astrophysical research.

\subsection{Convolutional Neural Networks}

DNNs, which are primarily implemented to extract features from grid-like data, including images and videos, are referred to as CNNs \citep{yamashita2018convolutional}. The architecture of a CNN comprises a sequence of discrete layers to generate an output from an input by means of differentiable transformations. CNNs are primarily powerful at automatically learning and improving filters through training \citep{galety2021deep}. They typically comprise the input layer, convolutional layers, pooling layers, and fully connected layers. Each layer plays a distinct role in the feature extraction and learning process \citep{li2021survey}. A CNN's fundamental building block is the convolutional layer. It uses learnable filters applied to the input data to derive spatial attributes. Different features of the input data are captured by stacking feature maps created via multiple filters \citep{aloysius2017review}. The convolution operation for functions $\rm f$ and $\rm g$ follows
\begin{equation}
    (\rm f * g)(t) = \int_{-\infty}^{\infty} f(\tau) g(t - \tau) \,\it d\tau,
\end{equation}

\noindent which aims to demonstrate how one function alters the form or structure of the other \citep{ajit2020review}.

To improve stability and reduce computational complexity, the pooling layer down-samples the feature maps by reducing their spatial dimensions. Max-pooling is a prevalent pooling method whereby the maximum value from each sub-matrix of the activation map is chosen to form a separate matrix \citep{jie2020runpool}. A pooling operation is defined as
\begin{align}
\rm x^{l}_{j} &= \rm f^l(u^{l}_{j}), \\
\rm u^{l}_{j} &= \rm \alpha^{l}_{j} \cdot f_{\text{pooling}}^{l}(x^{l-1}_{j}) + b^{l}_{j}.
\end{align}
In the given context, $\rm x^{l}_{j}$ and $\rm u^{\rm k}_{\rm j}$ respectively denote the output and net activation of the $\rm j$-th channel in the pooling layer $\rm l$. $\rm b^{l}_{j}$ represents the bias term for layer $\rm l$ and neuron $\rm j$. The function $\rm f^l(\cdot)$ represents the activation function connected to pooling layer $\rm l$, while $\rm \alpha^{l}_{j}$ denotes the weight coefficient associated with pooling layer $\rm l$. Additionally, $\rm f_{\text{pooling}}^{l}$ represents the pooling function applied in pooling layer $\rm l$. Together, these components describe the operations and transformations within the pooling layer of a neural network \citep{cong2023review}.

The fully connected layer is a conventional neural network layer in which each neuron in the preceding layer is connected to neurons in the currently active layer and is normally used at the end of the network to generate the final output \citep{krichen2023convolutional}. CNNs acquire optimal filters via backpropagation and gradient descent. The loss function is minimized by calculating the gradients of the loss with respect to the weights and iteratively updating the weights \citep{gu2018recent}. This method allows CNNs to autonomously extract features from data, rendering them exceptionally proficient at tasks related to 2D signals, including picture classification, object detection, and pattern recognition \citep{patil2021convolutional}. While CNNs excel in processing 2D data, they are not always suitable for 1D signals, especially when training data is limited or application-specific. To address this, 1D CNNs have been developed, which perform 1D convolutions instead of 2D \citep{kiranyaz20211d}. 1D CNNs have demonstrated remarkable performance in signal and audio processing. This effectiveness has also been leveraged in other domains, such as galaxy classification. Their simple and compact structure allows for economical, real-time hardware applications \citep{cavanagh2021morphological}. 

\subsection{Recurrent Neural Networks}

RNNs are a type of deep learning model developed for sequential data, rendering them particularly efficient for applications like time series analysis, natural language processing, and speech recognition \citep{schmidhuber2015deep}. RNNs maintain a persistent hidden state that captures information from previous time steps, allowing them to model temporal dependencies in sequential data \citep{yang2020lstm}. This distinctive property makes RNNs useful for situations where the sequence and context of data points are important. The fundamental concept of RNNs is the implementation of recurrent connections, allowing the network to manage sequences of different lengths \citep{sehovac2020deep}. 

Notwithstanding their advantages, RNNs encounter numerous problems such as vanishing and exploding gradients during training; gradients may either vanish (decrease excessively) or explode (increase excessively), which leads to complicated learning of long-term dependencies \citep{pascanu2013difficulty}. The additional issue is the challenge of training. Their sequential properties make them usually more difficult to train than feedforward networks. Besides, standard RNNs exhibit difficulties in retaining information across extended sequences, therefore constraining their efficacy in tasks necessitating long-term memory \citep{bengio1994learning}. Another challenge is their sequential processing, which results in lower computational efficiency, especially in long sequences \citep{prabowo2018lstm}.

In 1997, \citet{hochreiter1997long} presented the LSTM architecture to overcome conventional RNN limitations. LSTMs possess three primary gates to efficiently capture long-term dependence by including memory cells and gating systems to govern information flow. The input gate regulates the influx of new data into the memory cell, whilst the forget gate ascertains which information to eliminate from the memory cell \citep{nosouhian2021review}.  These gates for input $\rm x_{\rm t}$ are computed as follows
\begin{equation}
\begin{aligned}
\quad &\text{Input Gate: } \rm i_{\rm t} = \sigma(\rm W_{\rm i} \cdot [\rm h_{\rm t-1}, x_{\rm t}] + \rm b_{\rm i}), \\
\quad &\text{Forget Gate: } \rm f_{\rm t }= \sigma(\rm W_{\rm f} \cdot [\rm h_{\rm t-1}, x_{\rm t}] + \rm b_{\rm f}),\\
\quad &\text{Output Gate: } \rm o_{\rm t }= \sigma(\rm W_{\rm o} \cdot [\rm h_{\rm t-1}, x_{\rm t}] + \rm b_{\rm o}), 
\end{aligned}
\end{equation}
\noindent where ${\sigma}(.)$ is activation function, $\rm h_{\rm t-1}$ is hidden state at previous time step, $\rm W$ and $\rm b$ denote weight matrix and bias term, respectively \citep{van2020review}. The hidden state in an LSTM acts as the network's memory, summarizing information from previous time steps and enabling the model to retain and utilize long-term dependencies. The hidden state $\rm h_{\rm t}$ is then computed as
\begin{equation}
    \rm h_{\rm t} = \rm o_{\rm t} \odot \sigma(\rm C_{\rm t}),
\end{equation}
\noindent where $\rm o_{\rm t}$ is the output gate and $\rm C_{\rm t}$ is the cell state \citep{olah2015understanding}. The cell state is regulated by the forget and input gates, which selectively remove or add information. The cell state and candidate cell state $\tilde{\rm C}_{\rm t}$ are updated as follows
\begin{equation}
    \rm C_{\rm t} = \rm f_{\rm t} \odot \rm C_{\rm t-1} + \rm i_{\rm t} \odot \tilde{\rm C}_{\rm t},
    \tilde{C}_{\rm t} = \sigma(\rm W_{\rm C} \cdot [{h_{\rm t-1}}, \rm x_{\rm t}] + {b_{\rm C}}),
\end{equation}
\noindent where $\rm f_{\rm t}$ is the forget gate, $\rm i_{\rm t}$ is the input gate, $\odot$ denotes element-wise multiplication \citep{neil2016phased}.  

Gated Recurrent Units (GRUs) are a simpler alternative to LSTM that would minimize the complexity of computations while preserving performance \citep{chung2014empirical}. A single update gate is present in GRUs, which incorporates the input and forget gates. Additionally, a reset gate is introduced to regulate the flow of information. The update rule for the hidden state $\rm h_{\rm t}$ and candidate hidden state $\tilde{\rm h}_{\rm t}$ in a GRU is \citep{zargar2021introduction}
\begin{equation}
    \rm h_{\rm t} = (1 - \rm z_{\rm t}) \odot \rm h_{\rm t-1} + \rm z_{\rm t} \odot \tilde{\rm h}_{\rm t},
    \tilde{\rm h}_{\rm t} = \sigma(\rm W_{\rm h} (\rm r_{\rm t} \odot \rm h_{\rm t-1}) + \rm W_{\rm x} \rm x_{\rm t} + b),
\end{equation}

\noindent where $\rm z_{\rm t}$ is the update gate and $\rm r_{\rm t}$ denotes reset gate. The update and reset gate determine how much of the previous hidden state should be retained or ignored to compute the candidate state \citep{yiugit2021simple}
\begin{equation}
    {{r}_{\rm t}} = \sigma(\rm W_{\rm r} \cdot [{h_{\rm t-1}}, \rm x_{\rm t}] + 
    {b_{\rm r}}),
{{z}_{\rm t}} = \sigma(\rm W_{\rm z} \cdot {h_{\rm t-1}}, \rm x_{\rm t}] + {b_{\rm z}}).
\end{equation}
Both LSTMs and GRUs mitigate the vanishing gradient issue and enhance the capacity to capture long-term dependencies. Nevertheless, GRUs' simplified design and fewer number of parameters help to show more computational efficiency. For applications with limited processing resources, GRUs are rather suitable \citep{shewalkar2019performance}. These neural networks are potent instruments for modeling sequential data, with each architecture presenting distinct advantages and compromises. Their capacity to manage temporal relationships and analyze sequences of variable lengths has rendered them essential in domains such as natural language processing, time series analysis, and speech recognition.

\subsection{Stacked Generalization}

Combining several machine learning models into an ensemble will greatly raise prediction dependability, robustness, and accuracy. The human inclination to search out a variety of viewpoints prior to making critical decisions is the source of inspiration for this method \citep{ganaie2022ensemble}. Combining the predictions of several models referred to as base learners helps ensemble methods to outperform any single model on a given challenge \citep{ahrens2023pystacked}. Stacked generalization, sometimes known as stacking, is one such sophisticated ensemble method that combines the predictions of several base models using a secondary model termed a meta-model, hence improving predictive power and dependability \citep{healey2018mapping}. Introduced by \citet{wolpert1992stacked}, stacked generalization uses the predictions of individual models as inputs to a higher-level meta-model, so addressing the biases of individual models. Training base models on a subset of the data, then uses their predictions on the remaining data to train the meta-model. This enables the meta-model to learn how to maximize the strengths and minimize the shortcomings of the base models by means of an optimal combination of their forecasts. This leads to more precise and robust predictions \citep{bakurov2021genetic}.

\subsection{Hybrid Learning}

In machine learning, hybrid learning is the combination of several algorithms using their complementary capabilities to maximize general performance \citep{azevedo2024hybrid}. Combining CNNs and RNNs will help to use CNNs' spatial feature extraction powers and RNNs' sequential modeling capability \Citep{demiss2024application}. Recent studies, such as those conducted by \citep{shi2019hybrid,siraj2020hybrid,halbouni2022cnn}, demonstrate that hybrid CNN-RNN models are effective due to their proficiency in capturing both spatial and sequential information. Additionally, improved temporal modeling includes GRUs and LSTMs, which are both well-known types of RNNs. LSTMs are renowned for their ability to capture long-term dependencies, while GRUs offer a simpler architecture and faster training times \citep{fu2016using}. By integrating these architectures, hybrid models can achieve higher prediction accuracy and computational efficiency \citep{zeng2022parking}.

\section{Methodology}
\label{sec:4}
This section outlines the design and structure of our hybrid meta-model, which integrates LSTM and GRU networks with CNNs. The architecture is tailored to reconstruct the global 21-cm brightness temperature during the EoR. By combining the temporal modeling strengths of RNNs with the spatial feature extraction capabilities of CNNs, our model achieves robust and efficient predictions while significantly reducing computational costs compared to traditional simulation-based approaches.
    
\subsection{Dataset and Preprocessing}

In this work, we utilize the publicly accessible 21-cm signal during the EoR high-resolution modeling tool, the 21SSD simulation database (\href{https://21ssd.obspm.fr}{21ssd.obspm.fr}). Each simulation performed hydrodynamical processes within a 200 $\rm h^{-1}$ cMpc volume using $1024^3$ particles, resolving halos down to $\simeq 10^{10} \rm M_{\odot}$. These simulations consistently incorporate physical processes such as Wouthuysen-Field coupling, UV/X-ray heating, and star formation, making them invaluable for instrument calibration (e.g., SKA, HERA) and constraining parameters governing early galaxy formation. The 21SSD provides lightcone data at native and smoothed SKA-like resolutions, both with and without synthetic thermal noise. The dataset includes three-dimensional isotropic power spectra of $\delta \rm T_b$ fluctuations and redshift-dependent pixel distribution functions of $\delta \rm T_b$ values, with the latter highlighting non-Gaussian features critical for distinguishing astrophysical models. This dataset comprises 45 subsets spanning a comprehensive parameter space of astrophysical models, with X-ray efficiency $\rm f_X = \{0.1, 0.3, 1, 3, 10\}$, hard to soft X-ray ratio $\rm r_{H/S} = \{0, 0.5, 1\}$, and Lyman band emissivity efficiency $\rm f_\alpha = \{0.5, 1, 2\}$. Redshifts were treated as sequential features to enable time-series predictions, with each sample combining astrophysical parameters and redshift-dependent brightness temperatures. The partitioning systematically holds one parameter constant while varying the others, enabling clear analysis of each parameter's individual effects.

To reconstruct brightness temperatures, we focused on the probability density functions (PDFs) derived from lightcone data, which encode the 21-cm signal's evolution across 400 redshifts ($z = 6.00$-$14.97$). Each redshift bin contains 300 PDF values, but we selected the brightness temperature corresponding to the maximum PDF value per redshift, reducing the dataset from 5.4 million to 18,000 data points. Note that in smoothed particle hydrodynamics simulations, temperature estimates can occasionally deviate by a few Kelvins. A discrepancy is typically negligible in astrophysical contexts. However, when computing the 21-cm brightness temperature for very cold gas, such deviations can produce artificially strong absorption signals. For this reason, a lower cutoff at \( \rm T_b \simeq -200~\mathrm{mK} \) was introduced by \citet{semelin201721ssd} as a post-processing step to mitigate spurious extreme absorption features in the computed 21-cm brightness temperature. This choice was justified, as it prevented the average \( \rm T_b \) from being skewed by a few anomalously low values. However, this post-processing method introduces an artifact in the PDF with no physical meaning, manifesting as artificial spikes. Hence, we only include maximum PDFs corresponding to \( \rm T_b \) values below \(-190~\mathrm{mK}\) to correct an artifact due to a specific choice in the previous data handling. While reconstructing the EDGES global signal lies beyond this study's scope, we note that a comprehensive signal analysis method free of PDF-specific artifacts could enable direct comparison with or co-learning from the EDGES reported signal.

Standardizing feature scales helped to normalize them, thereby guaranteeing equal contribution during training. Features were turned to zero mean and unit variance using the Standard Scaler, improving model convergence and supporting anomaly detection \citep{rani2021decision,pedregosa2011scikit}. Preserving temporal order to prevent leaking, the dataset was split into training (70\%), test (30\%), and validation subsets (25\% of the training set), of which this stratification guarantees strong evaluation spanning several parameter combinations and redshift ranges. This ensures that the training process does not utilize all data points and promises an unbiased evaluation of the model's performance on unseen data. Providing a strong basis for training and validation, the final pre-processed data set strikes a balance between computational tractability and astrophysical diversity. The preprocessing pipeline reduces noise by separating the most likely brightness temperatures per redshift and normalizing characteristics. Important non-Gaussian signals necessary for parameter inference are preserved. This method guarantees effective generalization of the model over the complicated parameter space of the EoR. In this way, we ensure the model generalizes effectively across the EoR's parameter space. To conduct a more thorough examination of the influence of X-ray heating on brightness temperature, we expanded the range of $\rm f_X$ values to include sufficient values between 0.1 and 10. This process utilizes the original set of $\rm f_X$ values from the simulation. A core component of the algorithm generates mock data based on these simulations, ensuring that subsequent processing stages have access to these mock data to calculate residuals for final predictions of the global 21-cm brightness temperature.

\subsection{Feature Engineering}

In order to enhance the efficacy of machine learning models, feature engineering is an essential process. This process is used to transform raw data into meaningful data, which could be done in the form of creating new features, combining existing features, or selecting the most pertinent ones \citep{kuhn2019feature}. Correlation analysis helps to detect and remove redundant features by calculating correlations between features. This guarantees that the model includes the most pertinent features to avoid unnecessary complexity of the model \citep{blessie2012sigmis}. Additionally, correlation measures can evaluate the effectiveness of feature subsets by ensuring that features are highly correlated with the target variable but exhibit low correlation with each other \citep{yu2003feature}. We included a new feature obtained from the residuals between the forecasts of base models and the target values to improve the predictive capability of our model. The basis of this new feature is created by a DNN as the main base model, and applied RFs as auxiliary base models. A high correlation between features could negatively impact the model. For this reason, the feature-feature and feature-target correlations are calculated and visualized using a heatmap. By analyzing this heatmap, we ensured that the features did not exhibit high correlations. Ultimately, we applied the standard scaler to all features to improve model stability and performance. 

\begin{figure}
\resizebox{\hsize}{!}{\includegraphics{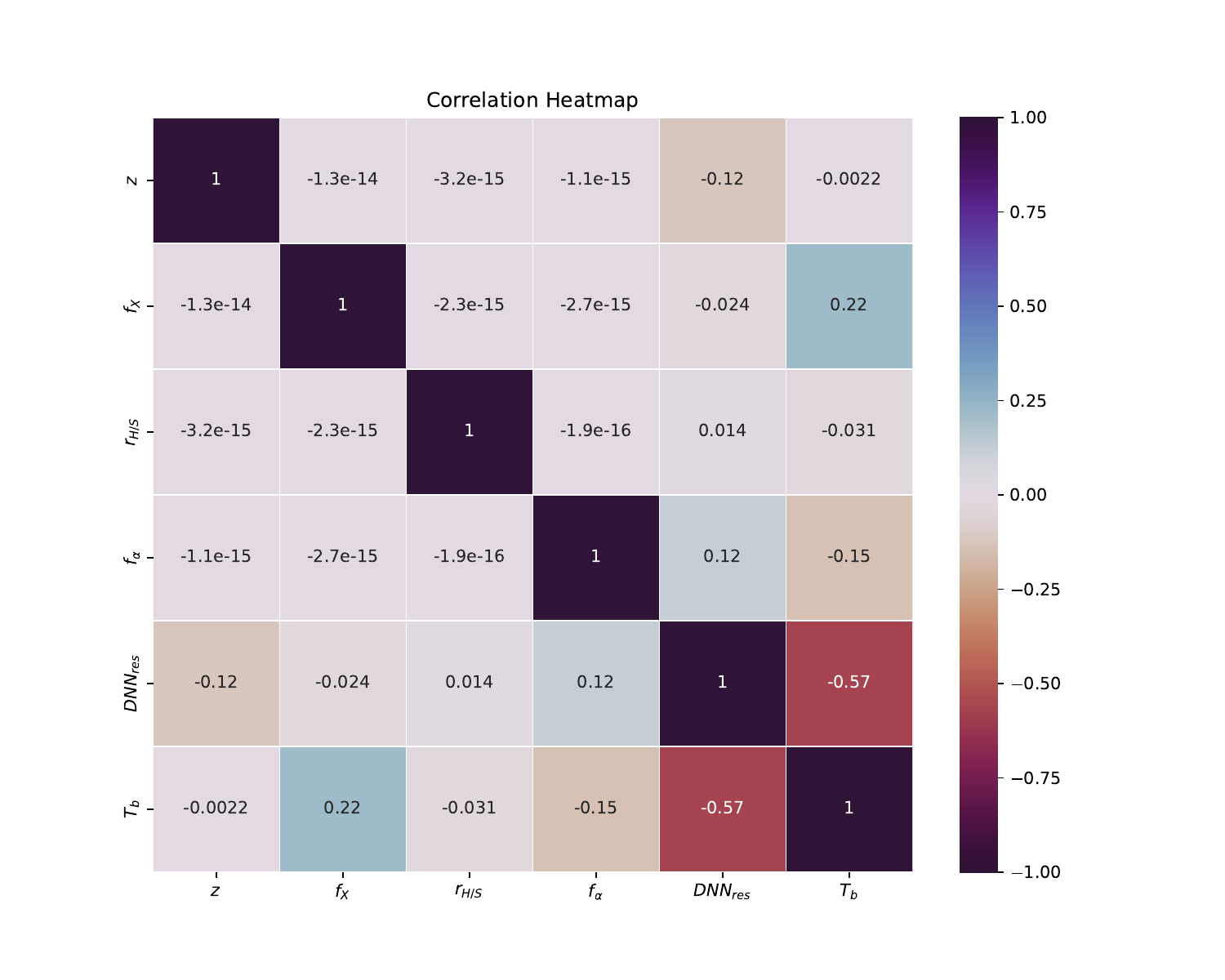}}
\caption{This heatmap shows feature-feature and feature-target correlation. The $\rm DNN_{res}$ feature shows strong correlations with $\rm f_X$ and $\rm T_b$, supporting its use in our model. All methods to prevent overfitting have been checked.}
\label{fig:1}
\end{figure}

Fig. \ref{fig:1} illustrates the feature-feature and feature-target correlation heatmap. This heatmap indicates negligible inter-feature correlations with a maximum of 0.12, which confirms there is no multicollinearity between predictors. Furthermore, our newly engineered feature shows substantially strong correlation with the target compared to baseline features. These results validate our feature selection methodology; the vanishingly small inter-feature correlations ensure predictor independence and model stability, while the pronounced feature-target relationships confirm the predictive relevance of our new feature set for brightness temperature reconstruction. The heatmap thus provides empirical justification for both our feature elimination criteria and the inclusion of newly derived feature.

\subsection{Bayesian Hyperparameter Tuning}

The hyperparameters are fundamental parameters that control the learning process and greatly affect model performance. The key hyperparameters in neural networks are the learning rate, the number of neurons in hidden layers, kernel sizes, dropout rates, and the regularizing parameters. In contrast to model parameters learned during training, the hyperparameters must be specified before training and demand careful adjustment to maximize performance \citep{victoria2021automatic}. There is no universal rule for choosing the hyperparameters since their ideal arrangement depends significantly on the particular dataset. Consequently, researchers often employ methodical testing in order to find the best configuration \citep{al2022multi}. Traditional methods for hyperparameter tuning, such as grid search and random search, are widely used but can be computationally expensive and inefficient. While grid search looks at all possible possibilities inside a particular search area, random search samples the hyperparameters randomly. None of these methods can optimize the search process using past performance statistics \citep{turner2021bayesian}. 
 The Bayesian optimization has become a strong and quick substitute for the hyperparameter tuning in order to overcome these restrictions. As opposed to grid or random search, Bayesian optimization is a probabilistic method that guides the search using past knowledge about the problem for ideal hyperparameters \citep{stuke2021efficient}. Bayesian optimization constructs a probabilistic model of the objective function, for instance, the validation loss, and uses it to identify the most promising hyperparameters for subsequent evaluation. This approach efficiently balances exploration and exploitation \citep{wang2023recent}. The optimization process is based on Bayes' Theorem as follows

\begin{equation}
\rm P(\Theta_M|D, M) = \frac{P(D|\Theta_M, M) \cdot P(\Theta_M|M)}{P(D|M)},
\end{equation}
\noindent where $\rm P(D|\Theta_M, M)$ is the likelihood, representing the probability of observing the data $\rm D$ given the hyperparameters $\rm \Theta_M$ and model $ \rm M $, $\rm P(\Theta_M|M)$ is the prior, representing the initial belief about the hyperparameters before observing the data. $\rm P(D|M)$ is the evidence, a normalization factor that ensures the posterior distribution integrates to 1, $\rm P(\Theta_M|D, M)$ denotes the posterior, representing the updated belief about the hyperparameters after observing the data \citep{verde2010statistical}.

 The Bayesian optimization improves its knowledge of the hyperparameter space by iteratively updating the posterior distribution and converging to the best configuration effectively \citep{shakya2022classification}. In this work, we tuned the hyperparameters of our neural network model using Bayesian optimization. The optimizer is set to investigate important hyperparameters, including the number of neurons, the kernel sizes, the dropout rates, the regularization parameters, and the learning rates. This method allows us to optimally search the high-dimensional hyperparameter space and find settings that optimize model performance. We obtained a more computationally efficient hyperparameter search by using the probabilistic framework of Bayesian optimization rather than conventional techniques.
 
\subsection{Architecture}
\label{Architecture}

\begin{figure*}
    \centering
    \includegraphics[width=15cm]{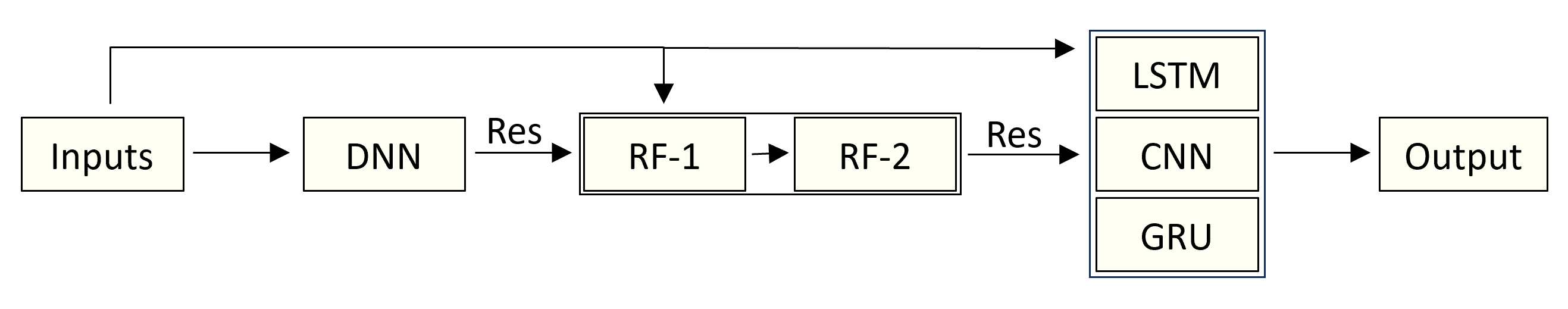}
    \caption{Proposed stacked hybrid architecture for reconstructing the global 21-cm brightness temperature from astrophysical parameters. The first stage combines a DNN and RFs to generate initial predictions and residuals. The second stage employs a hybrid recurrent-convolutional meta-model (LSTM, CNN, GRU) with proper regularization to leverage residuals and refine predictions through spatiotemporal pattern analysis.}
      \label{fig:2}
\end{figure*}

Our proposed architecture integrates deep learning and ensemble methods into a stacked hybrid framework (see Fig. \ref{fig:2} ) to reconstruct the brightness temperature from the astrophysical parameters. The system comprises two stages: a base model ensemble for preliminary predictions and feature extraction, followed by a hybrid recurrent-convolutional meta-model for prediction correction using residuals. The base ensemble consolidates a DNN and RFs to make use of their complementary strengths. The DNN employs dense layers with $\rm \tanh(x) = \frac{e^x - e^{-x}}{e^x + e^{-x}}$, $\rm \sigma(x) = \frac{1}{1 + e^{-x}}$, and $\rm ReLU(x) = \max(0, x)$ activations to model non-linear relationships. Furthermore, the batch normalization and the dropout contribute to stabilizing training and reducing overfitting. Parallel RFs with 100 estimators employ unpruned trees to manage noisy or incomplete data. The base models generate preliminary predictions and residuals, which are subsequently utilized to enhance the input space of the meta-model.
By use of a hybrid RNN-CNN architecture, the meta-model synthesizes temporal and spatial patterns. With L2 regularization ($1.00 \times 10^{-8}$) to limit weights, LSTM layers with 512 units process sequential dependencies using ReLU and sigmoid gating. Pairing L1 sparsity constraints ($4.67 \times 10^{-6}$) with $\tanh$ and ReLU activations, convolutional layers with 288 units extract localized spatial features with priority. Max pooling guarantees translational invariance using down-sampling. Enhanced by L2 regularization ($1.65 \times 10^{-5}$), GRUs with 512 units then simplify temporal dynamics with ReLU and softmax attention. Random deactivation of 0.15 of neurons follows the LSTM and GRU layers in the dropout layers.

The neural networks (DNN and meta-model) use the adaptive moment estimation (ADAM) optimizer \citep{kingma2014adam}, a stochastic optimization method that combines the advantages of AdaGrad \citep{duchi2011adaptive} and RMSProp \citep{ruder2016overview} for adaptive learning rates. ADAM efficiently computes parameter-specific updates by estimating first-order and second-order momentums of the gradients to optimize the training by minimizing the MSE loss. Early stopping activates after 10 epochs of validation loss stagnation. The meta-model incorporates raw astrophysical parameters alongside the new feature and iteratively refines predictions by resolving residuals. The architecture’s effectiveness stems from its stacked design. With RFs covering DNN's sensitivity to data sparsity, the DNN-RF ensemble diversifies feature representation. In addition to this, the combination of recurrent and convolutional procedures captures multi-scale relationships, which are essential for radiative process modeling of astrophysical systems. Even for incomplete or heterogeneous data, the framework provides a high-precision brightness temperature reconstruction by harmonizing ensemble variety.

We also evaluate the performance of four additional machine learning approaches: Support Vector Regression (SVR) with a radial basis function (RBF) kernel, an RNN with 512-unit LSTM layers, a 12-layer DNN trained for 150 epochs, and an RF regressor with 200 estimators. All models use the same train-test split ratio as the proposed model. SVR and RF were selected for their simplicity, as they require minimal hyperparameter tuning; however, RF is generally more resilient to overfitting. The LSTM-RNN is designed to capture temporal dependencies in sequential data, while the DNN efficiently models complex feature interactions. These comparisons aim to identify optimal base-learners that balance computational efficiency, predictive accuracy, and generalization capability.

\subsection{Metrics and Evaluations}
\label{evaluations}

It is essential to evaluate the performance of the machine learning models using a variety of metrics in order to ensure that the models are accurate and can be generalized to a wider population \citep{shanmugasundar2021comparative}. Two of the widely used metrics for this purpose are the MSE and the Mean Absolute Error (MAE). The MSE computes the mean squared difference between the predicted and the actual values, which makes it sensitive to outliers \citep{hodson2021mean}. For this reason, the MSE can be used as an indicator of the model's generalizing power on unseen data. Alternatively, the average absolute difference between the actual and expected values can be calculated using the MAE. This metric is a more accurate indicator of error magnitude \citep{nguyen2021influence}. The MSE and the MAE are described as

 \begin{equation}
     \rm MSE = \frac{1}{n} \sum_{i=1}^{n} (y_i - \hat{y}_i)^2,
 \end{equation}

    \begin{equation}
     \rm MAE = \frac{1}{n} \sum_{i=1}^{n} |y_i - \hat{y}_i|,
 \end{equation}

\noindent where $\rm n$ is the number of observations, $\rm y_i$ is the actual value, and $\rm \hat{y}_i$ is the predicted value. 

The coefficient of determination ($\rm R^2$) is another informative statistic that measures the fraction of variance in the dependent variable predictable from the independent factors. It ranges from 0 to 1; 1 marks ideal forecasts. From a mathematical standpoint \citep{chicco2021coefficient}

  \begin{equation}
      \rm R^2 = 1 - \frac{\sum_{i=1}^{n} (y_i - \hat{y}_i)^2}{\sum_{i=1}^{n} (y_i - \bar{y})^2},
  \end{equation}
    
  \noindent  where $\rm \bar{y}$ is the mean of the observed data.

We monitored these metrics to assess the accuracy and effectiveness of the model. Table \ref{Table:1} presents the performance of different algorithms on both the test and the training datasets. With an $\rm R^2$ score of 99.93\% and errors below 0.35 mK, our proposed model has excellent performance.

\begin{table*}[t]
\centering
\begin{tabular}{|c|c|c|c|c|c|c|}
\hline
\textbf{Method} & \textbf{MSE} & \textbf{MSE\textsubscript{test}} & \textbf{R\textsuperscript{2}score} & \textbf{R\textsuperscript{2}score\textsubscript{test}} & \textbf{MAE} & \textbf{MAE\textsubscript{test}} \\
\hline
DNN & 18.475& 17.154& 93.62\%& 93.91\%& 1.314& 1.313\\
RNN & 14.381& 11.989& 95.04\%& 95.74\%& 1.211& 1.186\\
SVR & 10.066& 7.936& 96.52\%& 97.18\%& 0.494& 0.531\\
RF & 1.955& 3.473& 99.31\%& 98.74\%& 0.198& 0.306\\
Proposed & 0.317& 0.324& 99.93\%& 99.91\%& 0.266& 0.272\\
\hline
\end{tabular}
\caption{Performance metrics for various methods on the overall and test datasets. The proposed model demonstrates superior performance, with minimal differences between training and test errors, indicating strong generalization.}
\label{Table:1}
\end{table*}

Whilst these models are able to perform better in different configurations, it is important to construct a robust model. One of the challenges in designing a model is excellent performance caused by overfitting. Overfitting occurs when a model performs well on the training set but poorly on unseen data. Many tactics could be used to help reduce the overfitting \citep{roelofs2019meta}. For example, random deactivation of neurons via dropout layers added during training promotes generalization. Moreover, using L1 and L2 regularization is an effective approach for penalizing large weights and encouraging simpler models \citep{ying2019overview}. Another method to prevent overfitting is early stopping, which requires separate validation sets during the training. The generalizing power of the model is found by evaluating the error in these validation sets along with the error of the training set in every epoch, then halting training when the validation loss ceases to improve \citep{tripathi2024extracting}.

\begin{figure*}
\centering
\includegraphics[width=15cm]{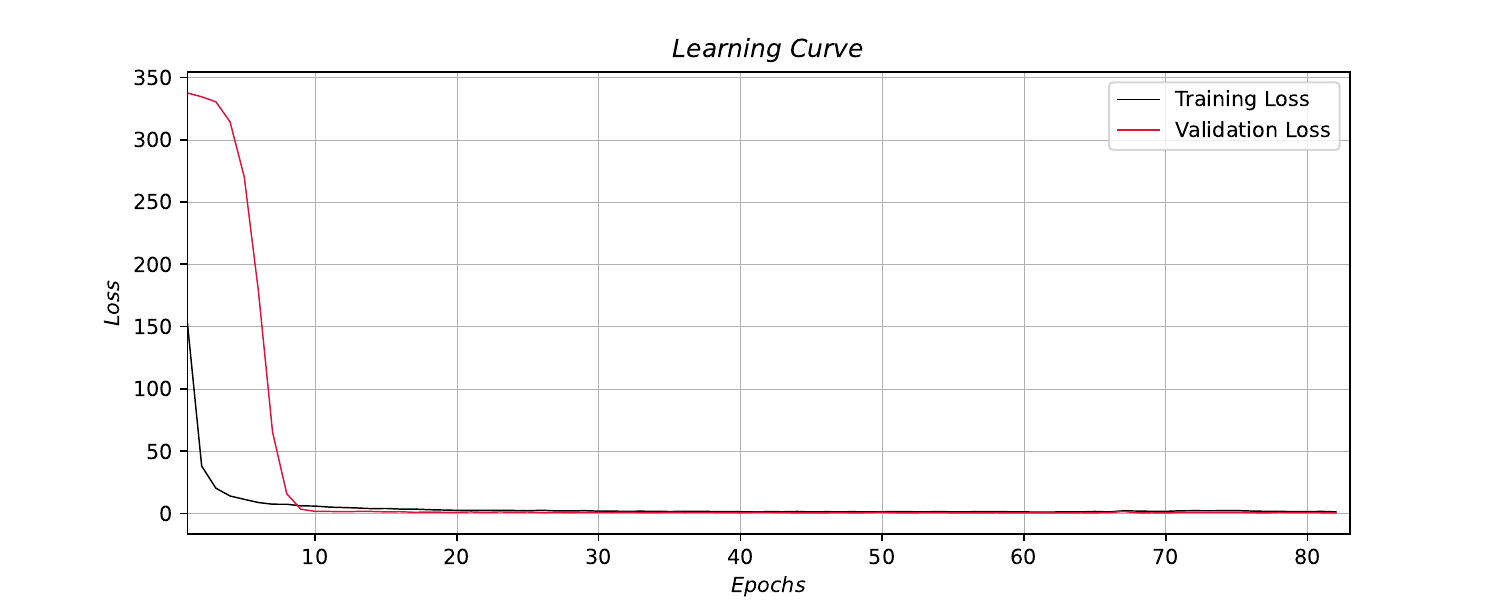}
\caption{Learning curve showing the training and validation losses over 82 epochs. Early stopping is employed to terminate the training process when no improvement in the validation loss is observed over 10 consecutive epochs. Both losses converge to near-zero values, indicating efficient learning and minimal overfitting.}
\label{fig:3}
\end{figure*}

To prevent overfitting in both base and meta models, we included dropout layers in addition to L1 and L2 regularization methods. To get a strong generalization for RFs, we let trees grow to their maximum depth and averaged their forecasts. Moreover, OOB estimation offers reliable projections of generalization error with no extra computational expenses. This estimation provides an unbiased assessment of the model's performance since it uses data points that are not included in the bootstrap samples. We split the dataset into the test and the training sets with a ratio of 3:7, and 25\% of the training set was randomly allocated as validation sets in each epoch to ensure the robustness of the model. The training and validation losses convergence over consecutive epochs is illustrated in the learning curve shown in Fig. \ref{fig:3}.

\section{Result}
 \label{sec:5}
\begin{figure*}[!t]
\centering
\includegraphics[width=17cm]{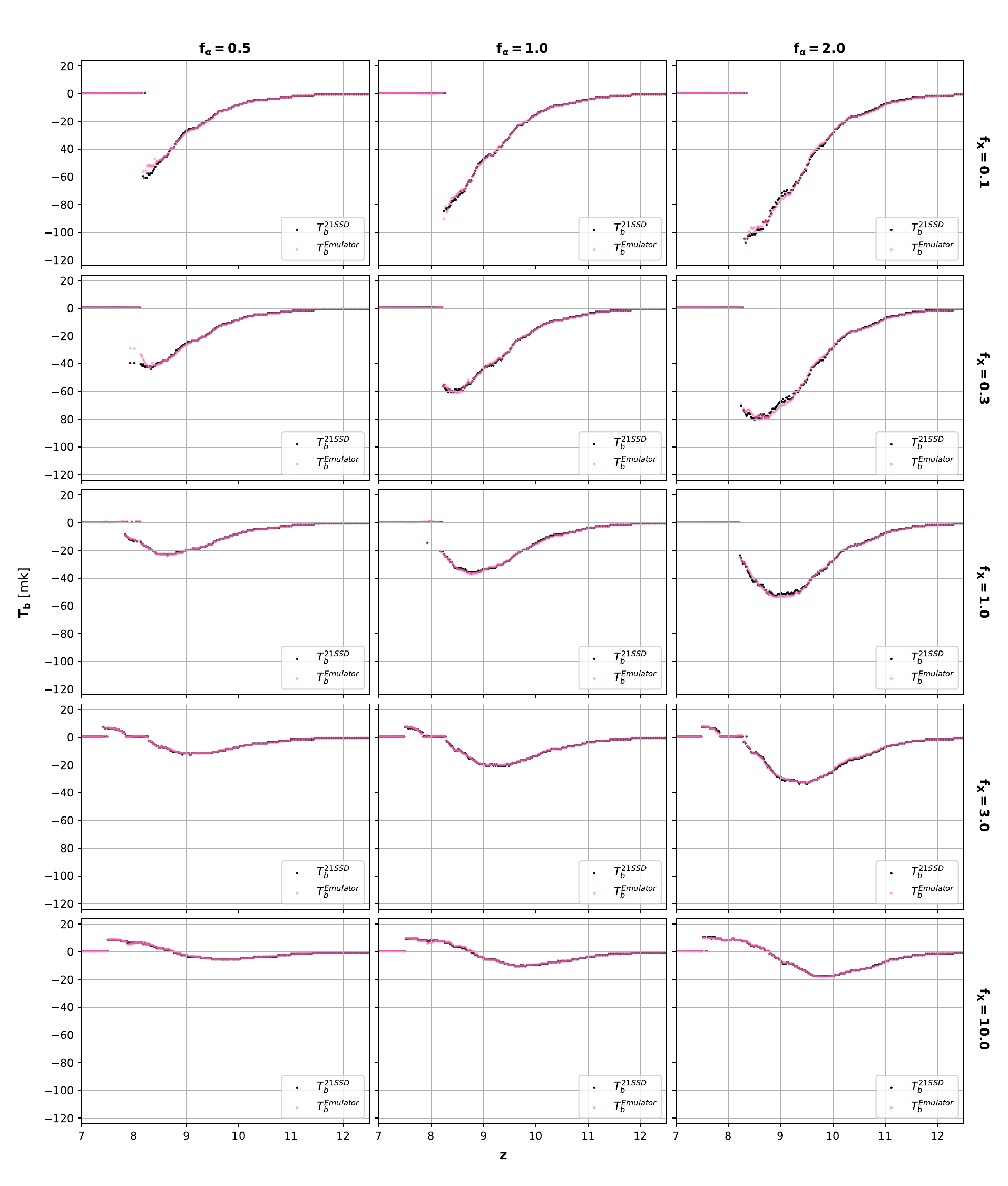}
\caption{The figure compares the global 21-cm brightness temperature from our emulator with the 21SSD simulation, where black points represent the original simulation data and pink points correspond to the reconstructed signal by the emulator. The architecture behind this method and the explanations are accessible in \S \ref{Architecture}. The emulator effectively tracks the data. Although the variables $\rm f_X$ and $\rm f_\alpha$ are changed within the range of models accessible from the simulations, the parameter $\rm r_{H/S}$ is fixed at zero. The emulator has struggled to capture the trend at higher levels of $\rm f_X$, leading to reduced accuracy, but this amount is not remarkable. Methods of evaluation are presented in \S \ref{evaluations}.}
\label{fig:4}
\end{figure*}

\begin{figure*}[!t]
\centering
\includegraphics[width=17cm]{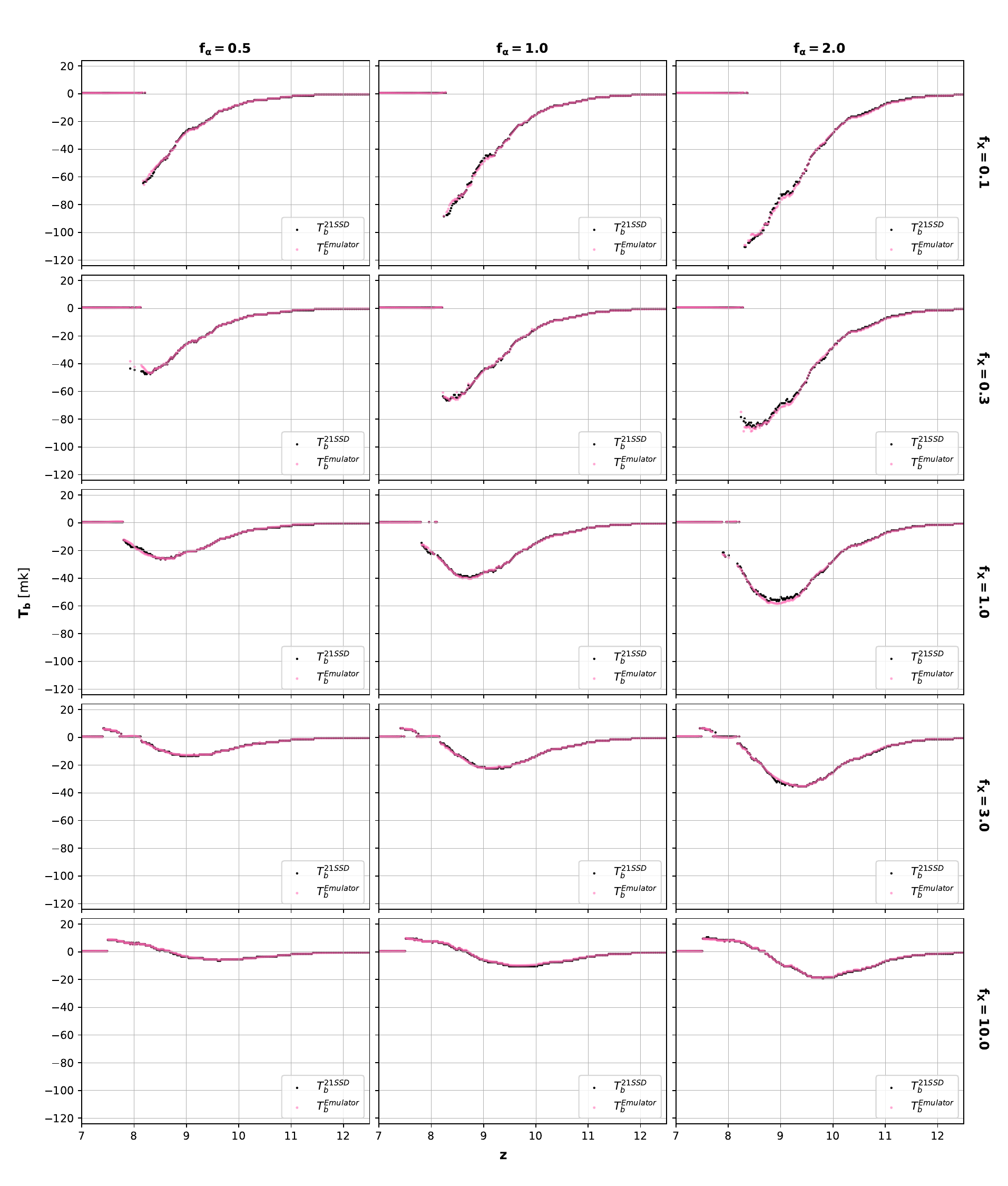}
\caption{The figure compares the global 21-cm brightness temperature from our emulator with the 21SSD simulation, where black points represent the original simulation data and pink points correspond to the reconstructed signal by the emulator. The architecture of this method and its explanations are available in \S \ref{Architecture}. The emulator performs acceptably to capture the trend. The variables $\rm f_X$ and $\rm f_\alpha$ change within the range of models derived from the simulations, and the parameter $\rm r_{H/S}$ remains constant at $0.5$. The emulator's fluctuations at $\rm f_X=1$ are inferior to those depicted in Fig.~\ref{fig:2}. The evaluation methods are detailed in \ref{evaluations}.}
\label{fig:5}
\end{figure*}

\begin{figure*}[!t]
\centering
\includegraphics[width=17cm]{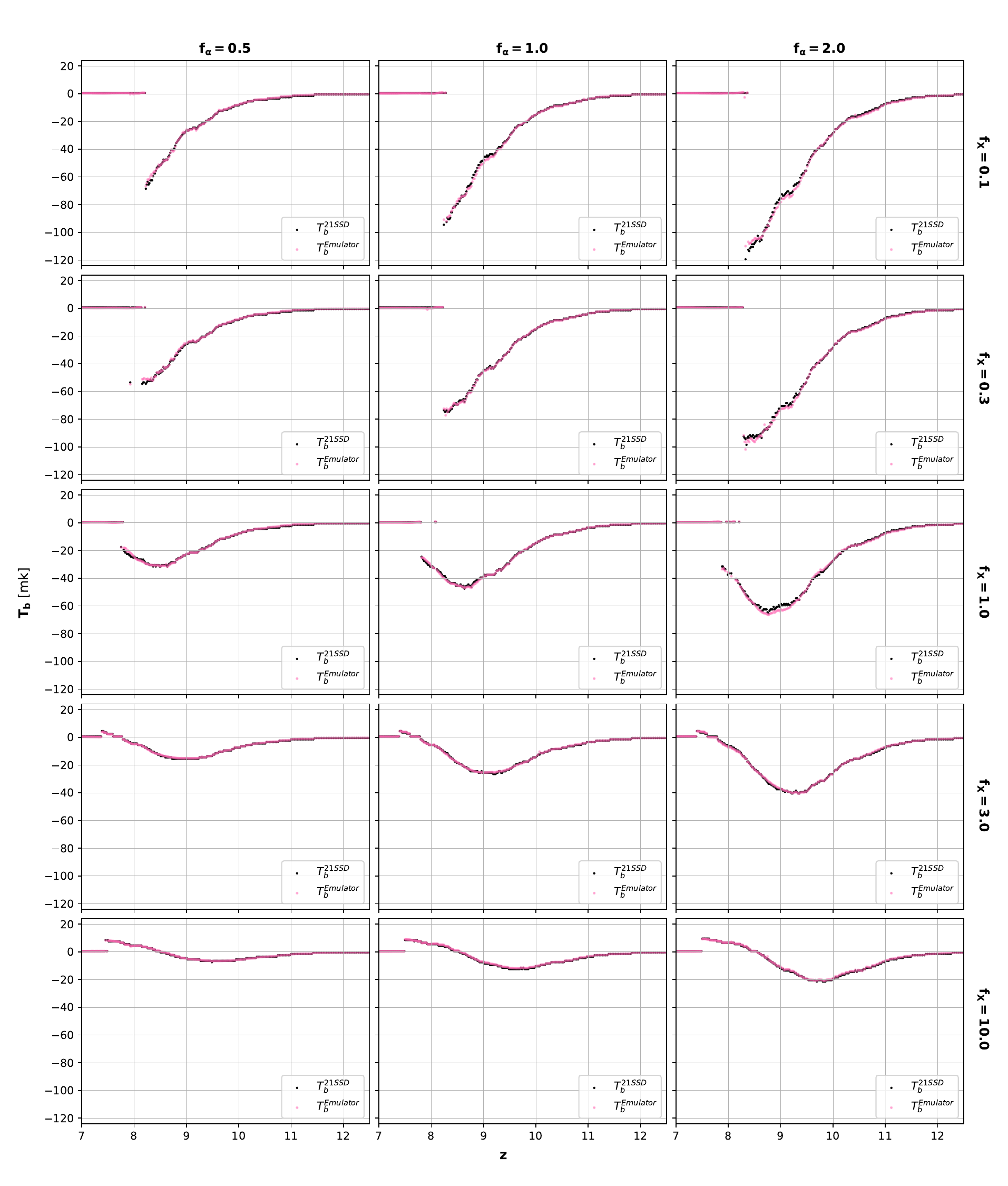}
\caption{The figure compares the global 21-cm brightness temperature from our emulator with the 21SSD simulation, where black points represent the original simulation data and pink points correspond to the reconstructed signal by the emulator. In \S \ref{Architecture}, one can discover the method's architecture and explanations. In terms of capturing the trend, the emulator succeeds adequately. Within the range of models obtained from the simulations, the variables $\rm f_X$ and $\rm f_\alpha$ change, whereas the parameter $\rm r_{H/S}$ holds constant at $1$. See \ref{evaluations} for a description of the evaluation procedures.}
\label{fig:6}
\end{figure*}

 As previously mentioned, we employ a hybrid stacked machine learning approach for reconstructing the global 21-cm brightness temperature span 45 astrophysical models with \(\rm f_X \in \{0.1, 0.3, 1, 3, 10\}\), \(\rm r_{H/S} \in \{0, 0.5, 1\}\), and \(\rm f_\alpha \in \{0.5, 1, 2\}\) (see §\ref{sec:2.3}). In this context, we compare our models with the simulation to validate our model. In Fig.~\ref{fig:4}, this comparison between the simulation and our model with fixed $\rm r_{H/S}$ to zero is shown, with black and pink points representing the original simulation data and the reconstructed signal, respectively. In Fig.~\ref{fig:5} $\rm r_{H/S}$ is fixed to $0.5$, and in Fig.~\ref{fig:6} $\rm r_{H/S}$ is fixed to $1$ with same analysis relative to Fig.~\ref{fig:4}. The emulator effectively tracks the data, although at higher amounts of $\rm f_\alpha$, the model exhibits a small amount of fluctuation relative to the simulation. The difference does not significantly change the overall accuracy of the model, as we can see the performance of our model at the loss figure per epoch in Fig.~\ref{fig:3}.

All methods mentioned in Table \ref{Table:1} were evaluated using training and testing splits to ensure fair comparison. It is evident that our proposed model demonstrates excellent performance with an $\rm R^2$ score of 99.93\% and errors below 0.35 mK. Notably, conventional approaches (DNN, RNN, SVR, RF) were optimized without dedicated validation sets, potentially risking hyperparameter overfitting to the test set. In contrast, our proposed method incorporated an independent validation phase during development to ensure robust generalization. The results demonstrate its substantial superiority: achieving near-perfect test scores ($\text{MSE}_{\text{test}} = 0.324$, $R^2_{\text{test}} = 99.91\%$, $\text{MAE}_{\text{test}} = 0.345$); outperforming even the strongest baseline (i.e. RF) by an order of magnitude in MSE reduction ($1.955 \rightarrow 0.311$) while maintaining exceptional consistency between training and testing metrics ($\Delta\text{MSE} = 0.007$). This stability is also evident in the learning curve (Fig.~\ref{fig:3}). There exist small spikes due to the stochastic nature of splitting and transient overfitting on challenging regions of the parameter space; they are below visually detectable thresholds, and the model quickly recovers; therefore, the final validation loss remains exceptionally low, which confirms robust overall performance. This contrasts with baselines like SVR, where test performance degraded significantly versus training ($\text{MSE}: 10.066 \rightarrow 7.936$), suggesting overfitting in conventional approaches.

As illustrated in Figs.~\ref{fig:4}--\ref{fig:6}, selecting the brightness temperature \( \rm T_b \) based on the maximum of its PDF at each redshift results in a step-like evolution. This behavior occurs because the PDF maxima tend to cluster around specific \( \rm T_b \) values over extended redshift ranges before abruptly transitioning to another preferred value. Thus, the step-like pattern is an inherent feature of this PDF-based selection method. In lower $\rm f_X$ values, delayed reionization results in deeper and sharper troughs, hence this step-like behavior of maximum PDFs reaches its maximum. While using the whole PDF instead of maxima could provide a fuller statistical picture, this approach is not adopted due to computational constraints; this might enhance the resolution of substructure in the signal at the cost of increased processing time.

A cornerstone of our analysis is extending parameter space coverage beyond the original 21SSD simulation's sparse sampling ($\rm f_X \in {0.1, 0.3, 1, 3, 10}$). Using hybrid stacked learning applied to simulation data and residuals between simulations and early-stage reconstructed data, we achieve two optimized sampling windows: a high-resolution window ($\rm f_X = 0.1$ to $1.0$ in steps of $0.1$) to resolve the nonlinear dependence of the 21-cm signal on faint, early X-ray sources, and a broad-scaling window ($\rm f_X = 1.0$ to $10$ in steps of $1.0$) to map saturation in X-ray heating efficiency. This expanded coverage enables unprecedented resolution in tracking the impact of X-ray heating on the global 21-cm signal morphology. 

\begin{figure*}[!t]
\centering
\includegraphics[width=17cm]{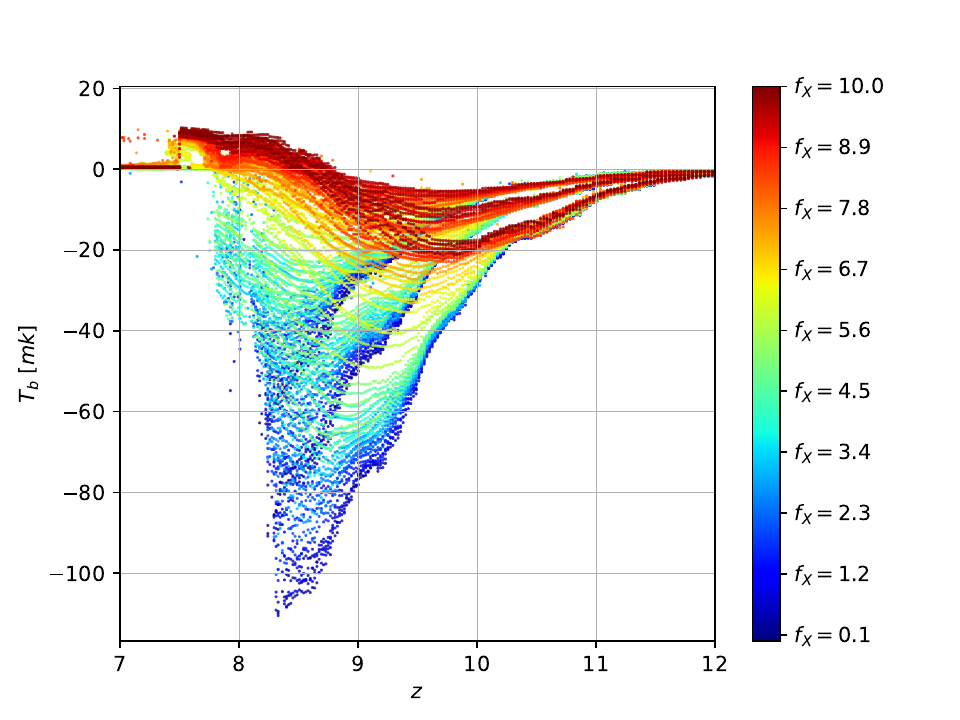}
\caption{The plot illustrates the global 21-cm} brightness temperature ($\rm T_b$) versus redshift ($z$) for various $\rm f_X$ values, with the color gradient representing $\rm f_X$ values ranging from 0.1 to 10.0. It demonstrates the effect of X-ray heating on the 21-cm signal, indicating that higher $\rm f_X$ values lead to more efficient X-ray heating, resulting in greater brightness temperatures at earlier redshifts. These are all of the models that are predicted using approaches that are outlined in \S \ref{sec:4}.
\label{fig:7}
\end{figure*}

The apparent reliance of the reionization timing on the X-ray efficiency parameter $\rm f_X$ is demonstrated by our machine learning models, which were tested against the 21SSD simulation. Plotting the findings shows a systematic trend in Fig.~\ref{fig:7}: lower $\rm f_X$ delays both the start and end points; higher $\rm f_X$ values correspond with an earlier onset and completion of reionization. For example, where the process happens much earlier in cosmic history, the reionization starts and ends at considerably greater redshifts compared to  $\rm f_X = 0.1$. This matter corresponds with the theoretical assumptions since increased X-ray efficiency accelerates the reionization process by promoting early ionization and the heating of the IGM.  The commencement and the completion of the reionization are gradually delayed as the values of $\rm f_X$ drop from 10.0 to 0.1. Consistent across all models is this inverse relationship between $\rm f_X$ and the redshift of reionization. This trend emphasizes the important part X-ray sources play in controlling the timeline of cosmic reionization. 

The IGM is heated to a temperature that exceeds the CMB when X-ray efficiency exceeds a binding threshold. The gas kinetic temperature $\rm T_k$ is coupled to the brightness temperature $\rm T_b$, which is determined by the spin temperature $\rm T_S$. At extremely high X-ray efficiencies ($ \geq 1$), $\rm T_k$ exceeds $\rm T_{CMB}$ by a significant margin, resulting in $\rm T_S\simeq T_k$ that can occur even when $\rm T_k < T_{CMB}$, provided $\rm x_{\mathrm{tot}} \gg1$. This results in the reversal of $\rm T_b$ from negative (absorption) to positive (emission). The threshold $\rm f_{\mathrm{X}}$ indicates the stage at which X-ray heating becomes dominant over the CMB and other radiative processes in setting the spin temperature $\rm T_{\mathrm{S}}$. At this point, efficient Ly$\alpha$ coupling ensures that $\rm T_{\mathrm{S}}$ is driven by the kinetic temperature $\rm T_{\mathrm{k}}$, which is elevated by X-ray heating, rather than by the CMB. Heating is inadequate to elevate $\rm T_k$ above $\rm T_{CMB}$ below this threshold; consequently, absorption takes precedence. Above it, the 21-cm signal is driven into positive territory, and absorption signatures are erased by rapid, uniform X-ray heating, generating large-scale emission regions. This transition indicates a phase transition in the thermal state of the IGM, which is determined by the predominance of X-ray heating over cosmic expansion and radiative cooling.
 
The soft X-rays from the AGNs surpass hard X-rays as the ratio $\rm r_{H/S}$ approaches 1. The IGM is uniformly heated by hard X-rays, which have a prolonged penetration depth and higher energy, thereby delaying or reducing localized heating. This uniform heating results in the preservation of more influential regions of gas cooler than the CMB, leading to a stronger 21-cm absorption (more negative brightness temperatures). In other words, as $\rm r_{H/S}$ decreases (i.e., the soft X-ray contribution increases), the overall 21-cm absorption signal becomes shallower. This is evident in the reduced depth of the troughs in Fig. \ref{fig:6} ($\rm r_{H/S}=1$) compared to Fig. \ref{fig:4} ($\rm r_{H/S}=0$). Conversely, soft X-rays from the AGNs produce localized heating, creating a patchwork of warmer (emission) regions and cooler (absorption) pockets. While this increases the spatial fragmentation of absorption zones, their individual depth is reduced, and they are embedded within emission-dominated regions, resulting in a net shallower absorption signal. The enhanced spatial complexity reflects the strong influence of soft X-ray heating.

The impact of the $\rm r_{H/S}$ decreases as the overall X-ray efficiency increases, as the IGM temperature is elevated more swiftly due to the enhanced heating, regardless of whether the X-rays are from the XRBs or the AGNs (Figs.~\ref{fig:4}--\ref{fig:6}). The 21-cm absorption signal falls (less negative brightness temperatures) as the IGM is heated to a temperature that approaches or exceeds that of the CMB at high X-ray efficiencies (Fig.~\ref{fig:7}, $\rm f_X \geq 3$). This overpowers the subtle variations in heating patterns caused by $\rm r_{H/S}$; the uniform heating of hard X-rays ($\rm r_{H/S} = 0$, Fig.~\ref{fig:4}) and the localized effects of soft X-rays ($\rm r_{H/S} = 1$, Fig.~\ref{fig:6}) become less distinguishable as the gas transitions to a warmer, emission-dominated phase (Fig.~\ref{fig:7}, $z < 10$). As a result, the brightness temperature structure is less influenced by the relative contribution of $\rm r_{H/S}$, as the primary factor is the total X-ray heating efficacy (Figs.~\ref{fig:4}--\ref{fig:6}, panels with $\rm f_X \geq 1$).

The Wouthuysen-Field effect is strengthened by the increased production of Ly$\alpha$ photons, which couples the spin temperature of the neutral hydrogen to the colder kinetic temperature of the IGM, as $\rm f_{\alpha}$ increases. It also raises the negative brightness temperature (more negative values) by broadening the 21-cm absorption, as the gas remains colder relative to the CMB. The onset of the reionization is also accelerated by a higher $\rm f_{\alpha}$, which increases the ionizing radiation from early stellar populations, thereby driving the earlier heating and ionization of neutral hydrogen. The coupling effect propagates more efficiently through the IGM, resulting in broader spatial and temporal regions of strong absorption (wider negative troughs) due to the increased emissivity. This phenomenon extends the phase during which cool, neutral gas dominates before reionization erodes it. Therefore, the 21-cm absorption signal is intensified, and its profundity and spatial/temporal extent are altered by an increase in $\rm f_{\alpha}$, which results in reionization occurring in earlier epochs. This leads to the emergence of three distinct branches, mostly visible at z = 9–11 as depicted in Fig.~\ref{fig:7}.

The 21-cm brightness temperature signal is influenced by the competing effects of $\rm f_{\alpha}$ and X-ray efficiency as they increase. The Ly$\alpha$ coupling is augmented by a higher $\rm f_{\alpha}$, which results in a deeper absorption trough and a closer proximity of the spin temperature to the cool gas kinetic temperature (more negative values). Nevertheless, the IGM temperature is elevated by the increased X-ray efficiency, which results from the enhanced heating, which counteracts this cooling. The gas warms toward or above the CMB temperature as X-ray heating becomes dominant, resulting in a reduction in the depth of the absorption signal (less negative brightness temperatures) and a shortening of the absorption phase. The rapid heating caused by high X-ray efficiency suppresses the effect of a more substantial $\rm f_{\alpha}$, which causes reionization to commence earlier and to widen the absorption trough temporally and spatially. This results in shallower absorption and an earlier transition to emission-dominated phases. The outcome is a "flattened" absorption signal, as reionization commences earlier due to $\rm f_{\alpha}$. However, the depth and width of the 21-cm absorption are reduced as X-rays heat the gas more rapidly, overriding the Ly$\alpha$ cooling influence.

Our machine learning system implements techniques such as regularization and cross-validation to enhance reliability and prevent overfitting. Training and validation were conducted on separate portions of the 21SSD simulated data. The model's performance indicators, such as the MSE and $\rm R^2$, demonstrate robust generalization to novel data. These methodologies ensured that the correlation between $\rm f_X$ and reionization accurately reflects physical trends rather than statistical anomalies, mitigating the risk of overfitting.

\section{Discussion and Conclusion}
\label{sec:6}

Our novel hybrid stacked machine learning architecture exhibits outstanding efficacy in reconstructing global 21-cm brightness temperature across the parameter space of $\rm f_X $, $\rm r_{H/S} $, and $\rm f_\alpha$. The reference 21SSD simulation requires $\sim$$3 \times 10^6$ CPU hours, whereas our emulator achieves significant computational efficiency, completing runs in approximately 15 minutes on a system with an Intel Core i7-6700K (8 threads), 24 GB RAM (2400 MHz). This represents a speedup of approximately six orders of magnitude. With an $\rm R^2$ score of 99.93\% with errors < 0.35 mK on both training and test datasets, the proposed model achieves exceptional accuracy, higher than baseline methods like DNNs, RNNs, SVR, and RFs (Table~\ref{Table:1}). While other presented methods exhibit shorter execution times, they fail to achieve the desired accuracy in the absence of strict regularization techniques. Implementation of several regularization techniques, including early stopping, dropout layers, L1/L2 penalties, and OOB estimations, confirms strong generalization by the small difference between training and test errors (e.g., MSE\textsubscript{test} = 0.324 mK vs. MSE = 0.317 mK).  Further underlining the absence of overfitting is the alignment of training and validation losses reaching minimal values (Fig.~\ref{fig:3}). Our method maintains conciseness and adaptability while incorporating multiple regularization techniques to effectively prevent overfitting and ensure reliable generalization across diverse test cases.

Our approach emphasizes the need for exact dataset separation and regularization.  We partitioned the data into training, test, and validation subsets to guarantee strong generalization.  Early stopping prevented more overfitting, dependent on validation loss plateauing. Growing trees to maximum depth and averaging forecasts boosted stability for ensemble techniques such as RFs; OOB estimation offered computationally efficient error bounds.

$\rm f_X$ and the redshift of reionization display an obvious inverse relationship in the analysis. Higher $\rm f_X$ values match earlier beginning and completion of reionization (Fig.~\ref{fig:7}), which is consistent with theoretical predictions that increased X-ray heating speeds ionization of the IGM. Slight fluctuations arise at modest $\rm f_X$ and extreme $\rm f_\alpha$ values, this pattern holds over all $\rm r_{H/S}$ configurations (Figs.~\ref{fig:4}--\ref{fig:6}).  For example, the emulator shows modest deviations from simulations at $\rm r_{H/S} = 0$ and $\rm f_X = 1$ (Fig.~\ref{fig:4}). Nonetheless, these differences fall within reasonable error margins ($\rm 0.35 mK$) and do not undermine the general predictive ability of the model.

The emulator's vulnerability to underrepresented parameter regimes in the 21SSD simulations is a paramount drawback.  For instance, the minor errors at $\rm f_X=1$ (Figs.~\ref{fig:4}--\ref{fig:6}) imply that adding adaptive sampling methods or extending the training dataset to include more general parameter ranges will help to increase resilience.  Furthermore, directly including physical restrictions (e.g., thermodynamic priors) into the loss function could help to reduce fluctuations in extreme regimes.

These findings have important consequences for the subsequent investigations of the 21-cm signal. Interpreting future data from equipment such as SKA depends on a fundamental observational baseline supplied by the proven inverse $\rm f_X$-redshift relationship. Moreover, the computational efficiency of our approach allows quick investigation of high-dimensional parameter spaces, bridging the gap between observational cosmology and simulation-based investigations.

At last, our work suggests machine learning as a powerful instrument for probing cosmic reionization.  Balancing accuracy, interpretability, and computing economy in the proposed hybrid model helps us to explore the complex interplay between astrophysical parameters and the EoR. This framework is intended to be extended from 21-cm signal prediction to a comprehensive tool applicable to any physical model, demonstrating its potential for broader applicability.

\begin{acknowledgements}
    Thanks to Prof. Geraint Lewis (University of Sydney) and Dr. B. Semelin (Observatoire de Paris) for their invaluable guidance and scientific expertise, which were essential to this work.
\end{acknowledgements}

  \bibliographystyle{aa} 
  \bibliography{main} 

\begin{thebibliography}{186}
\expandafter\ifx\csname natexlab\endcsname\relax\def\natexlab#1{#1}\fi

\bibitem[{Abel \& Haehnelt(1999)}]{abel1999radiative}
Abel, T. \& Haehnelt, M.~G. 1999, The Astrophysical Journal, 520, L13

\bibitem[{Adams {et~al.}(2023)Adams, Aguirre, Alexander, Ali, Baartman, Balfour, Barkana, Beardsley, Bernardi, Billings, {et~al.}}]{adams2023improved}
Adams, T., Aguirre, J.~E., Alexander, P., {et~al.} 2023, The Astrophysical Journal, 945, 124

\bibitem[{Aghanim {et~al.}(2020)Aghanim, Akrami, Ashdown, Aumont, Baccigalupi, Ballardini, Banday, Barreiro, Bartolo, Basak, {et~al.}}]{aghanim2020planck}
Aghanim, N., Akrami, Y., Ashdown, M., {et~al.} 2020, Astronomy \& Astrophysics, 641, A6

\bibitem[{Ahrens {et~al.}(2023)Ahrens, Hansen, \& Schaffer}]{ahrens2023pystacked}
Ahrens, A., Hansen, C.~B., \& Schaffer, M.~E. 2023, The Stata Journal, 23, 909

\bibitem[{Ajit {et~al.}(2020)Ajit, Acharya, \& Samanta}]{ajit2020review}
Ajit, A., Acharya, K., \& Samanta, A. 2020, in 2020 international conference on emerging trends in information technology and engineering (ic-ETITE), IEEE, 1--5

\bibitem[{Al~Bataineh {et~al.}(2022)Al~Bataineh, Kaur, \& Jalali}]{al2022multi}
Al~Bataineh, A., Kaur, D., \& Jalali, S. M.~J. 2022, IEEE Access, 10, 36963

\bibitem[{Ali {et~al.}(2015)Ali, Parsons, Zheng, Pober, Liu, Aguirre, Bradley, Bernardi, Carilli, Cheng, {et~al.}}]{ali201564}
Ali, Z.~S., Parsons, A.~R., Zheng, H., {et~al.} 2015, The Astrophysical Journal, 809, 61

\bibitem[{Aloysius \& Geetha(2017)}]{aloysius2017review}
Aloysius, N. \& Geetha, M. 2017, in 2017 international conference on communication and signal processing (ICCSP), IEEE, 0588--0592

\bibitem[{Alzubi {et~al.}(2018)Alzubi, Nayyar, \& Kumar}]{alzubi2018machine}
Alzubi, J., Nayyar, A., \& Kumar, A. 2018, in Journal of physics: conference series, Vol. 1142, IOP Publishing, 012012

\bibitem[{Amrani {et~al.}(2018)Amrani, Hamida, Liu, \& Langlois}]{amrani2018train}
Amrani, A., Hamida, A.~B., Liu, T., \& Langlois, O. 2018, in Transport Research Arena (TRA) 2018

\bibitem[{Azevedo {et~al.}(2024)Azevedo, Rocha, \& Pereira}]{azevedo2024hybrid}
Azevedo, B.~F., Rocha, A. M.~A., \& Pereira, A.~I. 2024, Machine Learning, 113, 4055

\bibitem[{Baek {et~al.}(2009)Baek, Di~Matteo, Semelin, Combes, \& Revaz}]{baek2009simulated}
Baek, S., Di~Matteo, P., Semelin, B., Combes, F., \& Revaz, Y. 2009, Astronomy \& Astrophysics, 495, 389

\bibitem[{Baek {et~al.}(2010)Baek, Semelin, Di~Matteo, Revaz, \& Combes}]{baek2010reionization}
Baek, S., Semelin, B., Di~Matteo, P., Revaz, Y., \& Combes, F. 2010, Astronomy \& Astrophysics, 523, A4

\bibitem[{Bakurov {et~al.}(2021)Bakurov, Castelli, Gau, Fontanella, \& Vanneschi}]{bakurov2021genetic}
Bakurov, I., Castelli, M., Gau, O., Fontanella, F., \& Vanneschi, L. 2021, Swarm and Evolutionary Computation, 65, 100913

\bibitem[{Barkana \& Loeb(2001)}]{barkana2001beginning}
Barkana, R. \& Loeb, A. 2001, Physics reports, 349, 125

\bibitem[{Barkana \& Loeb(2005)}]{barkana2005method}
Barkana, R. \& Loeb, A. 2005, The Astrophysical Journal, 624, L65

\bibitem[{Bengio {et~al.}(1994)Bengio, Simard, \& Frasconi}]{bengio1994learning}
Bengio, Y., Simard, P., \& Frasconi, P. 1994, IEEE transactions on neural networks, 5, 157

\bibitem[{Biau(2012)}]{biau2012analysis}
Biau, G. 2012, The Journal of Machine Learning Research, 13, 1063

\bibitem[{Blessie \& Karthikeyan(2012)}]{blessie2012sigmis}
Blessie, E.~C. \& Karthikeyan, E. 2012, Journal of Algorithms \& Computational Technology, 6, 385

\bibitem[{Bowman {et~al.}(2018)Bowman, Rogers, Monsalve, Mozdzen, \& Mahesh}]{bowman2018absorption}
Bowman, J.~D., Rogers, A.~E., Monsalve, R.~A., Mozdzen, T.~J., \& Mahesh, N. 2018, Nature, 555, 67

\bibitem[{Breiman(1996)}]{breiman1996out}
Breiman, L. 1996, University of California Berkeley

\bibitem[{Breitman {et~al.}(2024)Breitman, Mesinger, Murray, Prelogovi{\'c}, Qin, \& Trotta}]{breitman202421cmemu}
Breitman, D., Mesinger, A., Murray, S.~G., {et~al.} 2024, Monthly Notices of the Royal Astronomical Society, 527, 9833

\bibitem[{Cavanagh {et~al.}(2021)Cavanagh, Bekki, \& Groves}]{cavanagh2021morphological}
Cavanagh, M.~K., Bekki, K., \& Groves, B.~A. 2021, Monthly Notices of the Royal Astronomical Society, 506, 659

\bibitem[{Chakraborty \& Choudhury(2025)}]{chakraborty2025reionization}
Chakraborty, A. \& Choudhury, T.~R. 2025, arXiv preprint arXiv:2502.12004

\bibitem[{Chen {et~al.}(2023)Chen, Trac, Mukherjee, \& Cen}]{chen2023patchy}
Chen, N., Trac, H., Mukherjee, S., \& Cen, R. 2023, The Astrophysical Journal, 943, 138

\bibitem[{Cheng {et~al.}(2021)Cheng, Wu, Li, Yao, \& Min}]{cheng2021twd}
Cheng, S., Wu, Y., Li, Y., Yao, F., \& Min, F. 2021, Information Sciences, 579, 15

\bibitem[{Chicco {et~al.}(2021)Chicco, Warrens, \& Jurman}]{chicco2021coefficient}
Chicco, D., Warrens, M.~J., \& Jurman, G. 2021, PeerJ Computer Science, 7, e623

\bibitem[{Choi {et~al.}(2020)Choi, Coyner, Kalpathy-Cramer, Chiang, \& Campbell}]{choi2020introduction}
Choi, R.~Y., Coyner, A.~S., Kalpathy-Cramer, J., Chiang, M.~F., \& Campbell, J.~P. 2020, Translational vision science \& technology, 9, 14

\bibitem[{Chung {et~al.}(2014)Chung, Gulcehre, Cho, \& Bengio}]{chung2014empirical}
Chung, J., Gulcehre, C., Cho, K., \& Bengio, Y. 2014, arXiv preprint arXiv:1412.3555

\bibitem[{Cong \& Xiao(2014)}]{cong2014minimizing}
Cong, J. \& Xiao, B. 2014, in International conference on artificial neural networks, Springer, 281--290

\bibitem[{Cong \& Zhou(2023)}]{cong2023review}
Cong, S. \& Zhou, Y. 2023, Artificial Intelligence Review, 56, 1905

\bibitem[{Cutler {et~al.}(2012)Cutler, Cutler, \& Stevens}]{cutler2012random}
Cutler, A., Cutler, D.~R., \& Stevens, J.~R. 2012, Ensemble machine learning: Methods and applications, 157

\bibitem[{Datta {et~al.}(2016)Datta, Ghara, Majumdar, Choudhury, Bharadwaj, Roy, \& Datta}]{datta2016probing}
Datta, K.~K., Ghara, R., Majumdar, S., {et~al.} 2016, Journal of Astrophysics and Astronomy, 37, 1

\bibitem[{Dayal \& Ferrara(2018)}]{dayal2018early}
Dayal, P. \& Ferrara, A. 2018, Physics Reports, 780, 1

\bibitem[{de~Lera~Acedo(2019)}]{de2019reach}
de~Lera~Acedo, E. 2019, in 2019 International Conference on Electromagnetics in Advanced Applications (ICEAA), IEEE, 0626--0629

\bibitem[{Demiss \& Elsaigh(2024)}]{demiss2024application}
Demiss, B.~A. \& Elsaigh, W.~A. 2024, Engineering Research Express, 6, 032102

\bibitem[{Dewdney {et~al.}(2009)Dewdney, Hall, Schilizzi, \& Lazio}]{dewdney2009square}
Dewdney, P.~E., Hall, P.~J., Schilizzi, R.~T., \& Lazio, T. J.~L. 2009, Proceedings of the IEEE, 97, 1482

\bibitem[{Dhandha {et~al.}(2025)Dhandha, Gessey-Jones, Bevins, Pochinda, Fialkov, Tacchella, Acedo, Singh, \& Barkana}]{dhandha2025exploiting}
Dhandha, J., Gessey-Jones, T., Bevins, H.~T., {et~al.} 2025, arXiv preprint arXiv:2503.21687

\bibitem[{Duchi {et~al.}(2011)Duchi, Hazan, \& Singer}]{duchi2011adaptive}
Duchi, J., Hazan, E., \& Singer, Y. 2011, Journal of machine learning research, 12

\bibitem[{Dudek(2020)}]{dudek2020data}
Dudek, G. 2020, in 2020 International Joint Conference on Neural Networks (IJCNN), IEEE, 1--8

\bibitem[{Dutta {et~al.}(2018)Dutta, Jha, Sankaranarayanan, \& Tiwari}]{dutta2018output}
Dutta, S., Jha, S., Sankaranarayanan, S., \& Tiwari, A. 2018, in NASA Formal Methods Symposium, Springer, 121--138

\bibitem[{Dvorkin {et~al.}(2022)Dvorkin, Mishra-Sharma, Nord, Villar, Avestruz, Bechtol, {\'C}iprijanovi{\'c}, Connolly, Garrison, Narayan, {et~al.}}]{dvorkin2022machine}
Dvorkin, C., Mishra-Sharma, S., Nord, B., {et~al.} 2022, arXiv preprint arXiv:2203.08056

\bibitem[{Eide {et~al.}(2018)Eide, Graziani, Ciardi, Feng, Kakiichi, \& Di~Matteo}]{eide2018epoch}
Eide, M.~B., Graziani, L., Ciardi, B., {et~al.} 2018, Monthly Notices of the Royal Astronomical Society, 476, 1174

\bibitem[{Erkal(2015)}]{erkal2015investigating}
Erkal, D. 2015, Monthly Notices of the Royal Astronomical Society, 451, 904

\bibitem[{Ewen \& Purcell(1951)}]{ewen1951observation}
Ewen, H.~I. \& Purcell, E.~M. 1951, Nature, 168, 356

\bibitem[{Faaique(2024)}]{faaique2024overview}
Faaique, M. 2024, International Journal of Mathematics, Statistics, and Computer Science, 2, 96

\bibitem[{Fialkov {et~al.}(2014)Fialkov, Barkana, \& Visbal}]{fialkov2014observable}
Fialkov, A., Barkana, R., \& Visbal, E. 2014, Nature, 506, 197

\bibitem[{Fialkov {et~al.}(2017)Fialkov, Cohen, Barkana, \& Silk}]{fialkov2017constraining}
Fialkov, A., Cohen, A., Barkana, R., \& Silk, J. 2017, Monthly Notices of the Royal Astronomical Society, 464, 3498

\bibitem[{Field(1958)}]{field1958excitation}
Field, G.~B. 1958, Proceedings of the IRE, 46, 240

\bibitem[{Field(1959{\natexlab{a}})}]{field1959attempt}
Field, G.~B. 1959{\natexlab{a}}, Astrophysical Journal, vol. 129, p. 525, 129, 525

\bibitem[{Field(1959{\natexlab{b}})}]{field1959spin}
Field, G.~B. 1959{\natexlab{b}}, Astrophysical Journal, vol. 129, p. 536, 129, 536

\bibitem[{Finkelstein {et~al.}(2015)Finkelstein, Ryan, Papovich, Dickinson, Song, Somerville, Ferguson, Salmon, Giavalisco, Koekemoer, {et~al.}}]{finkelstein2015evolution}
Finkelstein, S.~L., Ryan, R.~E., Papovich, C., {et~al.} 2015, The Astrophysical Journal, 810, 71

\bibitem[{Fu {et~al.}(2016)Fu, Zhang, \& Li}]{fu2016using}
Fu, R., Zhang, Z., \& Li, L. 2016, in 2016 31st Youth academic annual conference of Chinese association of automation (YAC), IEEE, 324--328

\bibitem[{Furlanetto \& Oh(2016)}]{furlanetto2016reionization}
Furlanetto, S.~R. \& Oh, S.~P. 2016, Monthly Notices of the Royal Astronomical Society, 457, 1813

\bibitem[{Furlanetto {et~al.}(2006)Furlanetto, Oh, \& Briggs}]{furlanetto2006cosmology}
Furlanetto, S.~R., Oh, S.~P., \& Briggs, F.~H. 2006, Physics reports, 433, 181

\bibitem[{Galety {et~al.}(2021)Galety, Mukthar, Husham, Maaroof, \& Rofoo}]{galety2021deep}
Galety, M.~G., Mukthar, A., Husham, F., Maaroof, R.~J., \& Rofoo, F. 2021, Technium, 3

\bibitem[{Ganaie {et~al.}(2022)Ganaie, Hu, Malik, Tanveer, \& Suganthan}]{ganaie2022ensemble}
Ganaie, M.~A., Hu, M., Malik, A.~K., Tanveer, M., \& Suganthan, P.~N. 2022, Engineering Applications of Artificial Intelligence, 115, 105151

\bibitem[{Genuer {et~al.}(2017)Genuer, Poggi, Tuleau-Malot, \& Villa-Vialaneix}]{genuer2017random}
Genuer, R., Poggi, J.-M., Tuleau-Malot, C., \& Villa-Vialaneix, N. 2017, Big Data Research, 9, 28

\bibitem[{Glazer {et~al.}(2018)Glazer, Rau, \& Trac}]{glazer2018reionization}
Glazer, D., Rau, M.~M., \& Trac, H. 2018, arXiv preprint arXiv:1808.00553

\bibitem[{Gnedin \& Madau(2022)}]{gnedin2022modeling}
Gnedin, N.~Y. \& Madau, P. 2022, Living Reviews in Computational Astrophysics, 8, 3

\bibitem[{Greig \& Mesinger(2015)}]{greig201521cmmc}
Greig, B. \& Mesinger, A. 2015, Monthly Notices of the Royal Astronomical Society, 449, 4246

\bibitem[{Grossberg(2013)}]{grossberg2013recurrent}
Grossberg, S. 2013, Scholarpedia, 8, 1888

\bibitem[{Gu {et~al.}(2018)Gu, Wang, Kuen, Ma, Shahroudy, Shuai, Liu, Wang, Wang, Cai, {et~al.}}]{gu2018recent}
Gu, J., Wang, Z., Kuen, J., {et~al.} 2018, Pattern recognition, 77, 354

\bibitem[{Guo {et~al.}(2024)Guo, Fang, Feng, \& Zhang}]{guo2024multi}
Guo, X., Fang, G., Feng, H., \& Zhang, R. 2024, Research in Astronomy and Astrophysics, 24, 125019

\bibitem[{Halbouni {et~al.}(2022)Halbouni, Gunawan, Habaebi, Halbouni, Kartiwi, \& Ahmad}]{halbouni2022cnn}
Halbouni, A., Gunawan, T.~S., Habaebi, M.~H., {et~al.} 2022, IEEE Access, 10, 99837

\bibitem[{Hasan {et~al.}(2016)Hasan, Nasser, Ahmad, \& Molla}]{hasan2016feature}
Hasan, M. A.~M., Nasser, M., Ahmad, S., \& Molla, K.~I. 2016, Journal of information security, 7, 129

\bibitem[{Healey {et~al.}(2018)Healey, Cohen, Yang, Brewer, Brooks, Gorelick, Hernandez, Huang, Hughes, Kennedy, {et~al.}}]{healey2018mapping}
Healey, S.~P., Cohen, W.~B., Yang, Z., {et~al.} 2018, Remote Sensing of Environment, 204, 717

\bibitem[{{HERA Collaboration}(2023)}]{hera2023first}
{HERA Collaboration}. 2023, The Astrophysical Journal, 945, 124

\bibitem[{Hirata(2006)}]{hirata2006wouthuysen}
Hirata, C.~M. 2006, Monthly Notices of the Royal Astronomical Society, 367, 259

\bibitem[{Hochreiter \& Schmidhuber(1997)}]{hochreiter1997long}
Hochreiter, S. \& Schmidhuber, J. 1997, Neural computation, 9, 1735

\bibitem[{Hodson {et~al.}(2021)Hodson, Over, \& Foks}]{hodson2021mean}
Hodson, T.~O., Over, T.~M., \& Foks, S.~S. 2021, Journal of Advances in Modeling Earth Systems, 13, e2021MS002681

\bibitem[{Hogg \& Foreman-Mackey(2018)}]{hogg2018data}
Hogg, D.~W. \& Foreman-Mackey, D. 2018, The Astrophysical Journal Supplement Series, 236, 11

\bibitem[{Hosseini {et~al.}(2023)Hosseini, Salmasi, Tabasi, \& Firouzjaee}]{hosseini2023new}
Hosseini, S.~M., Salmasi, B.~S., Tabasi, S.~S., \& Firouzjaee, J.~T. 2023, arXiv preprint arXiv:2306.12954

\bibitem[{Jaiswal \& Samikannu(2017)}]{jaiswal2017application}
Jaiswal, J.~K. \& Samikannu, R. 2017, in 2017 world congress on computing and communication technologies (WCCCT), Ieee, 65--68

\bibitem[{Jamieson {et~al.}(2024)Jamieson, Smith, Neyer, Kannan, Garaldi, Vogelsberger, Hernquist, Zier, Shen, \& Kakiichi}]{jamieson2024thesan}
Jamieson, N., Smith, A., Neyer, M., {et~al.} 2024, arXiv preprint arXiv:2411.08943

\bibitem[{Janitza {et~al.}(2018)Janitza, Celik, \& Boulesteix}]{janitza2018computationally}
Janitza, S., Celik, E., \& Boulesteix, A.-L. 2018, Advances in Data Analysis and Classification, 12, 885

\bibitem[{Jie \& Wanda(2020)}]{jie2020runpool}
Jie, H.~J. \& Wanda, P. 2020, International Journal of Computational Intelligence Systems, 13, 66

\bibitem[{Kacprzak \& Fluri(2022)}]{kacprzak2022deeplss}
Kacprzak, T. \& Fluri, J. 2022, Physical Review X, 12, 031029

\bibitem[{Keller {et~al.}(2023)Keller, Nikolic, Thyagarajan, Carilli, Bernardi, Charles, Bester, Smirnov, Kern, Dillon, {et~al.}}]{keller2023search}
Keller, P.~M., Nikolic, B., Thyagarajan, N., {et~al.} 2023, Monthly Notices of the Royal Astronomical Society, 524, 583

\bibitem[{Kingma \& Ba(2014)}]{kingma2014adam}
Kingma, D.~P. \& Ba, J. 2014, arXiv preprint arXiv:1412.6980

\bibitem[{Kiranyaz {et~al.}(2021)Kiranyaz, Avci, Abdeljaber, Ince, Gabbouj, \& Inman}]{kiranyaz20211d}
Kiranyaz, S., Avci, O., Abdeljaber, O., {et~al.} 2021, Mechanical systems and signal processing, 151, 107398

\bibitem[{Kosowsky(2003)}]{kosowsky2003atacama}
Kosowsky, A. 2003, New Astronomy Reviews, 47, 939

\bibitem[{Kravtsov {et~al.}(1997)Kravtsov, Klypin, \& Khokhlov}]{kravtsov1997adaptive}
Kravtsov, A.~V., Klypin, A.~A., \& Khokhlov, A.~M. 1997, The Astrophysical Journal Supplement Series, 111, 73

\bibitem[{Krichen(2023)}]{krichen2023convolutional}
Krichen, M. 2023, Computers, 12, 151

\bibitem[{Kuhn \& Johnson(2019)}]{kuhn2019feature}
Kuhn, M. \& Johnson, K. 2019, Feature engineering and selection: A practical approach for predictive models (Chapman and Hall/CRC)

\bibitem[{Li {et~al.}(2021)Li, Liu, Yang, Peng, \& Zhou}]{li2021survey}
Li, Z., Liu, F., Yang, W., Peng, S., \& Zhou, J. 2021, IEEE transactions on neural networks and learning systems, 33, 6999

\bibitem[{Liu {et~al.}(2016)Liu, Slatyer, \& Zavala}]{liu2016contributions}
Liu, H., Slatyer, T.~R., \& Zavala, J. 2016, Physical Review D, 94, 063507

\bibitem[{Lomba \& H{\o}ye(2014)}]{lomba2014critical}
Lomba, E. \& H{\o}ye, J.~S. 2014, Molecular Physics, 112, 2892

\bibitem[{Lui \& Wolf(2019)}]{lui2019construction}
Lui, H.~F. \& Wolf, W.~R. 2019, Journal of Fluid Mechanics, 872, 963

\bibitem[{Ma {et~al.}(2022)Ma, Fiaschi, Ciardi, Busch, \& Eide}]{ma2022crash}
Ma, Q.-B., Fiaschi, S., Ciardi, B., Busch, P., \& Eide, M.~B. 2022, Monthly Notices of the Royal Astronomical Society, 513, 1513

\bibitem[{Madau {et~al.}(2004)Madau, Rees, Volonteri, Haardt, \& Oh}]{madau2004early}
Madau, P., Rees, M.~J., Volonteri, M., Haardt, F., \& Oh, S.~P. 2004, The Astrophysical Journal, 604, 484

\bibitem[{Mapelli {et~al.}(2006)Mapelli, Ferrara, \& Pierpaoli}]{mapelli2006impact}
Mapelli, M., Ferrara, A., \& Pierpaoli, E. 2006, Monthly Notices of the Royal Astronomical Society, 369, 1719

\bibitem[{Mertens {et~al.}(2024)Mertens, Bobin, \& Carucci}]{mertens2024retrieving}
Mertens, F.~G., Bobin, J., \& Carucci, I.~P. 2024, Monthly Notices of the Royal Astronomical Society, 527, 3517

\bibitem[{Mesinger {et~al.}(2011)Mesinger, Furlanetto, \& Cen}]{mesinger201121cmfast}
Mesinger, A., Furlanetto, S., \& Cen, R. 2011, Monthly Notices of the Royal Astronomical Society, 411, 955

\bibitem[{Mesinger {et~al.}(2016)Mesinger, Greig, \& Sobacchi}]{mesinger2016evolution}
Mesinger, A., Greig, B., \& Sobacchi, E. 2016, Monthly Notices of the Royal Astronomical Society, 459, 2342

\bibitem[{Mirabel {et~al.}(2011)Mirabel, Dijkstra, Laurent, Loeb, \& Pritchard}]{mirabel2011stellar}
Mirabel, I., Dijkstra, M., Laurent, P., Loeb, A., \& Pritchard, J. 2011, Astronomy \& Astrophysics, 528, A149

\bibitem[{Mirocha {et~al.}(2017)Mirocha, Harker, \& Burns}]{mirocha2017global}
Mirocha, J., Harker, G. J.~A., \& Burns, J.~O. 2017, The Astrophysical Journal, 843, 46

\bibitem[{Mitra {et~al.}(2015)Mitra, Choudhury, \& Ferrara}]{mitra2015cosmic}
Mitra, S., Choudhury, T.~R., \& Ferrara, A. 2015, Monthly Notices of the Royal Astronomical Society: Letters, 454, L76

\bibitem[{Mittal \& Kulkarni(2022)}]{mittal2022first}
Mittal, S. \& Kulkarni, G. 2022, Monthly Notices of the Royal Astronomical Society, 515, 3947

\bibitem[{Mondal \& Barkana(2023)}]{mondal2023prospects}
Mondal, R. \& Barkana, R. 2023, Nature Astronomy, 7, 1025

\bibitem[{Morales \& Wyithe(2010)}]{morales2010reionization}
Morales, M.~F. \& Wyithe, J. S.~B. 2010, Annual review of astronomy and astrophysics, 48, 127

\bibitem[{Murray {et~al.}(2020)Murray, Greig, Mesinger, Mu{\~n}oz, Qin, Park, \& Watkinson}]{murray202021cmfast}
Murray, S.~G., Greig, B., Mesinger, A., {et~al.} 2020, arXiv preprint arXiv:2010.15121

\bibitem[{Nasir \& D’Aloisio(2020)}]{nasir2020observing}
Nasir, F. \& D’Aloisio, A. 2020, Monthly Notices of the Royal Astronomical Society, 494, 3080

\bibitem[{Nasteski(2017)}]{nasteski2017overview}
Nasteski, V. 2017, Horizons. b, 4, 56

\bibitem[{Neben {et~al.}(2016)Neben, Bradley, Hewitt, DeBoer, Parsons, Aguirre, Ali, Cheng, Ewall-Wice, Patra, {et~al.}}]{neben2016hydrogen}
Neben, A.~R., Bradley, R.~F., Hewitt, J.~N., {et~al.} 2016, The Astrophysical Journal, 826, 199

\bibitem[{Neil {et~al.}(2016)Neil, Pfeiffer, \& Liu}]{neil2016phased}
Neil, D., Pfeiffer, M., \& Liu, S.-C. 2016, Advances in neural information processing systems, 29

\bibitem[{Nguyen {et~al.}(2021)Nguyen, Ly, Ho, Al-Ansari, Le, Tran, Prakash, \& Pham}]{nguyen2021influence}
Nguyen, Q.~H., Ly, H.-B., Ho, L.~S., {et~al.} 2021, Mathematical Problems in Engineering, 2021, 4832864

\bibitem[{Ni {et~al.}(2023)Ni, Li, \& Zhang}]{ni2023cmb}
Ni, S., Li, Y., \& Zhang, X. 2023, arXiv preprint arXiv:2310.07358

\bibitem[{Nosouhian {et~al.}(2021)Nosouhian, Nosouhian, \& Khoshouei}]{nosouhian2021review}
Nosouhian, S., Nosouhian, F., \& Khoshouei, A.~K. 2021, Preprints

\bibitem[{Ntampaka {et~al.}(2019)Ntampaka, Avestruz, Boada, Caldeira, Cisewski-Kehe, Di~Stefano, Dvorkin, Evrard, Farahi, Finkbeiner, {et~al.}}]{ntampaka2019role}
Ntampaka, M., Avestruz, C., Boada, S., {et~al.} 2019, arXiv preprint arXiv:1902.10159

\bibitem[{Olah(2015)}]{olah2015understanding}
Olah, C. 2015, Understanding LSTM Networks

\bibitem[{Paciga {et~al.}(2013)Paciga, Albert, Bandura, Chang, Gupta, Hirata, Odegova, Pen, Peterson, Roy, {et~al.}}]{paciga2013simulation}
Paciga, G., Albert, J.~G., Bandura, K., {et~al.} 2013, Monthly Notices of the Royal Astronomical Society, 433, 639

\bibitem[{Pascanu {et~al.}(2013)Pascanu, Mikolov, \& Bengio}]{pascanu2013difficulty}
Pascanu, R., Mikolov, T., \& Bengio, Y. 2013, in International conference on machine learning, Pmlr, 1310--1318

\bibitem[{Patil \& Rane(2021)}]{patil2021convolutional}
Patil, A. \& Rane, M. 2021, Information and Communication Technology for Intelligent Systems: Proceedings of ICTIS 2020, Volume 1, 21

\bibitem[{Patil {et~al.}(2025)Patil, {\v{S}}oltinsk{\`y}, Maitra, \& Kulkarni}]{patil2025efficient}
Patil, S.~K., {\v{S}}oltinsk{\`y}, T., Maitra, S., \& Kulkarni, G. 2025, arXiv preprint arXiv:2507.11611

\bibitem[{Patra {et~al.}(2013)Patra, Subrahmanyan, Raghunathan, \& Udaya~Shankar}]{patra2013saras}
Patra, N., Subrahmanyan, R., Raghunathan, A., \& Udaya~Shankar, N. 2013, Experimental Astronomy, 36, 319

\bibitem[{Pedregosa {et~al.}(2011)Pedregosa, Varoquaux, Gramfort, Michel, Thirion, Grisel, Blondel, Prettenhofer, Weiss, Dubourg, {et~al.}}]{pedregosa2011scikit}
Pedregosa, F., Varoquaux, G., Gramfort, A., {et~al.} 2011, Journal of machine learning research, 12, 2825

\bibitem[{Prabowo {et~al.}(2018)Prabowo, Warnars, Budiharto, Kistijantoro, Heryadi, {et~al.}}]{prabowo2018lstm}
Prabowo, Y.~D., Warnars, H. L. H.~S., Budiharto, W., {et~al.} 2018, in 2018 Indonesian association for pattern recognition international conference (INAPR), IEEE, 51--56

\bibitem[{Pritchard \& Furlanetto(2007)}]{pritchard2007descending}
Pritchard, J.~R. \& Furlanetto, S.~R. 2007, Monthly Notices of the Royal Astronomical Society, 376, 1680

\bibitem[{Pritchard \& Loeb(2012)}]{pritchard201221}
Pritchard, J.~R. \& Loeb, A. 2012, Reports on Progress in Physics, 75, 086901

\bibitem[{Purwono {et~al.}(2022)Purwono, Ma'arif, Rahmaniar, Fathurrahman, Frisky, \& ul~Haq}]{purwono2022understanding}
Purwono, P., Ma'arif, A., Rahmaniar, W., {et~al.} 2022, International Journal of Robotics and Control Systems, 2, 739

\bibitem[{Rani {et~al.}(2021)Rani, Kumar, Ahmed, \& Jain}]{rani2021decision}
Rani, P., Kumar, R., Ahmed, N. M.~S., \& Jain, A. 2021, Journal of Reliable Intelligent Environments, 7, 263

\bibitem[{Resende \& Drummond(2018)}]{resende2018survey}
Resende, P. A.~A. \& Drummond, A.~C. 2018, ACM Computing Surveys (CSUR), 51, 1

\bibitem[{Ricotti(2002)}]{ricotti2002did}
Ricotti, M. 2002, Monthly Notices of the Royal Astronomical Society, 336, L33

\bibitem[{Roelofs {et~al.}(2019)Roelofs, Shankar, Recht, Fridovich-Keil, Hardt, Miller, \& Schmidt}]{roelofs2019meta}
Roelofs, R., Shankar, V., Recht, B., {et~al.} 2019, Advances in neural information processing systems, 32

\bibitem[{Rozos {et~al.}(2021)Rozos, Dimitriadis, Mazi, \& Koussis}]{rozos2021multilayer}
Rozos, E., Dimitriadis, P., Mazi, K., \& Koussis, A.~D. 2021, Hydrology, 8, 67

\bibitem[{Ruder(2016)}]{ruder2016overview}
Ruder, S. 2016, arXiv preprint arXiv:1609.04747

\bibitem[{Ruhl {et~al.}(2004)Ruhl, Ade, Carlstrom, Cho, Crawford, Dobbs, Greer, Holzapfel, Lanting, Lee, {et~al.}}]{ruhl2004south}
Ruhl, J., Ade, P.~A., Carlstrom, J.~E., {et~al.} 2004, in Millimeter and Submillimeter Detectors for Astronomy II, Vol. 5498, SPIE, 11--29

\bibitem[{Rybicki \& Lightman(2024)}]{rybicki2024radiative}
Rybicki, G.~B. \& Lightman, A.~P. 2024, Radiative processes in astrophysics (John Wiley \& Sons)

\bibitem[{Santos {et~al.}(2010)Santos, Ferramacho, Silva, Amblard, \& Cooray}]{santos2010fast}
Santos, M.~G., Ferramacho, L., Silva, M., Amblard, A., \& Cooray, A. 2010, Monthly Notices of the Royal Astronomical Society, 406, 2421

\bibitem[{Saxena {et~al.}(2023)Saxena, Cole, Gazagnes, Meerburg, Weniger, \& Witte}]{saxena2023constraining}
Saxena, A., Cole, A., Gazagnes, S., {et~al.} 2023, Monthly Notices of the Royal Astronomical Society, 525, 6097

\bibitem[{Schmidhuber(2015)}]{schmidhuber2015deep}
Schmidhuber, J. 2015, Neural networks, 61, 85

\bibitem[{Schmit \& Pritchard(2018)}]{schmit2018emulation}
Schmit, C.~J. \& Pritchard, J.~R. 2018, Monthly Notices of the Royal Astronomical Society, 475, 1213

\bibitem[{Scott \& Rees(1990)}]{scott199021}
Scott, D. \& Rees, M.~J. 1990, Monthly Notices of the Royal Astronomical Society, vol. 247, p. 510, 247, 510

\bibitem[{Sehovac \& Grolinger(2020)}]{sehovac2020deep}
Sehovac, L. \& Grolinger, K. 2020, Ieee Access, 8, 36411

\bibitem[{Semelin {et~al.}(2007)Semelin, Combes, \& Baek}]{semelin2007lyman}
Semelin, B., Combes, F., \& Baek, S. 2007, Astronomy \& Astrophysics, 474, 365

\bibitem[{Semelin {et~al.}(2017)Semelin, Eames, Bolgar, \& Caillat}]{semelin201721ssd}
Semelin, B., Eames, E., Bolgar, F., \& Caillat, M. 2017, Monthly Notices of the Royal Astronomical Society, 472, 4508

\bibitem[{Shakya {et~al.}(2022)Shakya, Biswas, \& Pal}]{shakya2022classification}
Shakya, A., Biswas, M., \& Pal, M. 2022, in Radar Remote Sensing (Elsevier), 175--186

\bibitem[{Shanmugasundar {et~al.}(2021)Shanmugasundar, Vanitha, {\v{C}}ep, Kumar, Kalita, \& Ramachandran}]{shanmugasundar2021comparative}
Shanmugasundar, G., Vanitha, M., {\v{C}}ep, R., {et~al.} 2021, Processes, 9, 2015

\bibitem[{Shao {et~al.}(2023)Shao, Xu, Wang, Yang, Li, Zhang, \& Chen}]{shao202321}
Shao, Y., Xu, Y., Wang, Y., {et~al.} 2023, Nature Astronomy, 7, 1116

\bibitem[{Shewalkar {et~al.}(2019)Shewalkar, Nyavanandi, \& Ludwig}]{shewalkar2019performance}
Shewalkar, A., Nyavanandi, D., \& Ludwig, S.~A. 2019, Journal of Artificial Intelligence and Soft Computing Research, 9, 235

\bibitem[{Shi {et~al.}(2019)Shi, Wang, Wang, Liu, \& Yan}]{shi2019hybrid}
Shi, X., Wang, T., Wang, L., Liu, H., \& Yan, N. 2019, in 2019 Asia-Pacific signal and information processing association annual summit and conference (APSIPA ASC), IEEE, 939--944

\bibitem[{Shimabukuro {et~al.}(2023)Shimabukuro, Hasegawa, Kuchinomachi, Yajima, \& Yoshiura}]{shimabukuro2023exploring}
Shimabukuro, H., Hasegawa, K., Kuchinomachi, A., Yajima, H., \& Yoshiura, S. 2023, Publications of the Astronomical Society of Japan, 75, S1

\bibitem[{Shimabukuro \& Semelin(2017)}]{shimabukuro2017analysing}
Shimabukuro, H. \& Semelin, B. 2017, Monthly Notices of the Royal Astronomical Society, 468, 3869

\bibitem[{Shukla {et~al.}(2016)Shukla, Mellema, Iliev, \& Shapiro}]{shukla2016effects}
Shukla, H., Mellema, G., Iliev, I.~T., \& Shapiro, P.~R. 2016, Monthly Notices of the Royal Astronomical Society, 458, 135

\bibitem[{Singh {et~al.}(2022)Singh, Nambissan~T, Subrahmanyan, Udaya~Shankar, Girish, Raghunathan, Somashekar, Srivani, \& Sathyanarayana~Rao}]{singh2022detection}
Singh, S., Nambissan~T, J., Subrahmanyan, R., {et~al.} 2022, Nature Astronomy, 6, 607

\bibitem[{Siraj \& Ahad(2020)}]{siraj2020hybrid}
Siraj, M.~S. \& Ahad, M. 2020, in 2020 Joint 9th International Conference on Informatics, Electronics \& Vision (ICIEV) and 2020 4th International Conference on Imaging, Vision \& Pattern Recognition (icIVPR), IEEE, 1--7

\bibitem[{Smagulova \& James(2019)}]{smagulova2019survey}
Smagulova, K. \& James, A.~P. 2019, The European Physical Journal Special Topics, 228, 2313

\bibitem[{{\v{S}}oltinsk{\`y} {et~al.}(2025){\v{S}}oltinsk{\`y}, Kulkarni, Tendulkar, \& Bolton}]{vsoltinsky2025prospects}
{\v{S}}oltinsk{\`y}, T., Kulkarni, G., Tendulkar, S.~P., \& Bolton, J.~S. 2025, Monthly Notices of the Royal Astronomical Society, 537, 364

\bibitem[{Speiser {et~al.}(2019)Speiser, Miller, Tooze, \& Ip}]{speiser2019comparison}
Speiser, J.~L., Miller, M.~E., Tooze, J., \& Ip, E. 2019, Expert systems with applications, 134, 93

\bibitem[{Stuke {et~al.}(2021)Stuke, Rinke, \& Todorovi{\'c}}]{stuke2021efficient}
Stuke, A., Rinke, P., \& Todorovi{\'c}, M. 2021, Machine Learning: Science and Technology, 2, 035022

\bibitem[{Sunyaev \& Zel'Dovich(1980)}]{sunyaev1980microwave}
Sunyaev, R. \& Zel'Dovich, Y.~B. 1980, Annual review of astronomy and astrophysics, 18, 537

\bibitem[{Thakur \& Konde(2021)}]{thakur2021fundamentals}
Thakur, A. \& Konde, A. 2021, International Journal for Research in Applied Science and Engineering Technology, 9, 407

\bibitem[{Tilvi {et~al.}(2020)Tilvi, Malhotra, Rhoads, Coughlin, Zheng, Finkelstein, Veilleux, Mobasher, Wang, Probst, {et~al.}}]{tilvi2020onset}
Tilvi, V., Malhotra, S., Rhoads, J., {et~al.} 2020, The Astrophysical Journal Letters, 891, L10

\bibitem[{Tingay {et~al.}(2013)Tingay, Goeke, Bowman, Emrich, Ord, Mitchell, Morales, Booler, Crosse, Wayth, {et~al.}}]{tingay2013murchison}
Tingay, S.~J., Goeke, R., Bowman, J.~D., {et~al.} 2013, Publications of the Astronomical Society of Australia, 30, e007

\bibitem[{Tripathi {et~al.}(2024)Tripathi, Datta, Choudhury, \& Majumdar}]{tripathi2024extracting}
Tripathi, A., Datta, A., Choudhury, M., \& Majumdar, S. 2024, Monthly Notices of the Royal Astronomical Society, 528, 1945

\bibitem[{Turner {et~al.}(2021)Turner, Eriksson, McCourt, Kiili, Laaksonen, Xu, \& Guyon}]{turner2021bayesian}
Turner, R., Eriksson, D., McCourt, M., {et~al.} 2021, in NeurIPS 2020 Competition and Demonstration Track, PMLR, 3--26

\bibitem[{Van~de Hulst(1945)}]{van1945radiogolven}
Van~de Hulst, H. 1945, Nederlandsch Tijdschrift voor Natuurkunde, 11, 210

\bibitem[{Van~de Schoot {et~al.}(2021)Van~de Schoot, Depaoli, King, Kramer, M{\"a}rtens, Tadesse, Vannucci, Gelman, Veen, Willemsen, {et~al.}}]{van2021bayesian}
Van~de Schoot, R., Depaoli, S., King, R., {et~al.} 2021, Nature Reviews Methods Primers, 1, 1

\bibitem[{van Haarlem {et~al.}(2013)van Haarlem, Wise, Gunst, Heald, McKean, Hessels, de~Bruyn, Nijboer, Swinbank, Fallows, {et~al.}}]{van2013lofar}
van Haarlem, M.~P., Wise, M.~W., Gunst, A., {et~al.} 2013, Astronomy \& astrophysics, 556, A2

\bibitem[{Van~Houdt {et~al.}(2020)Van~Houdt, Mosquera, \& N{\'a}poles}]{van2020review}
Van~Houdt, G., Mosquera, C., \& N{\'a}poles, G. 2020, Artificial Intelligence Review, 53, 5929

\bibitem[{Varsamopoulos {et~al.}(2017)Varsamopoulos, Criger, \& Bertels}]{varsamopoulos2017decoding}
Varsamopoulos, S., Criger, B., \& Bertels, K. 2017, Quantum Science and Technology, 3, 015004

\bibitem[{Vazza {et~al.}(2019)Vazza, Ettori, Roncarelli, Angelinelli, Br{\"u}ggen, \& Gheller}]{vazza2019detecting}
Vazza, F., Ettori, S., Roncarelli, M., {et~al.} 2019, Astronomy \& Astrophysics, 627, A5

\bibitem[{Venkatesan {et~al.}(2001)Venkatesan, Giroux, \& Shull}]{venkatesan2001heating}
Venkatesan, A., Giroux, M.~L., \& Shull, J.~M. 2001, The Astrophysical Journal, 563, 1

\bibitem[{Venumadhav {et~al.}(2018)Venumadhav, Dai, Kaurov, \& Zaldarriaga}]{venumadhav2018heating}
Venumadhav, T., Dai, L., Kaurov, A., \& Zaldarriaga, M. 2018, Physical Review D, 98, 103513

\bibitem[{Verde(2010)}]{verde2010statistical}
Verde, L. 2010, in Lectures on Cosmology: Accelerated Expansion of the Universe (Springer), 147--177

\bibitem[{Victoria \& Maragatham(2021)}]{victoria2021automatic}
Victoria, A.~H. \& Maragatham, G. 2021, Evolving Systems, 12, 217

\bibitem[{Vonlanthen {et~al.}(2011)Vonlanthen, Semelin, Baek, \& Revaz}]{vonlanthen2011distinctive}
Vonlanthen, P., Semelin, B., Baek, S., \& Revaz, Y. 2011, Astronomy \& Astrophysics, 532, A97

\bibitem[{Wan {et~al.}(2017)Wan, Dahlsten, Kristj{\'a}nsson, Gardner, \& Kim}]{wan2017quantum}
Wan, K.~H., Dahlsten, O., Kristj{\'a}nsson, H., Gardner, R., \& Kim, M. 2017, npj Quantum information, 3, 36

\bibitem[{Wang {et~al.}(2023)Wang, Jin, Schmitt, \& Olhofer}]{wang2023recent}
Wang, X., Jin, Y., Schmitt, S., \& Olhofer, M. 2023, ACM Computing Surveys, 55, 1

\bibitem[{Wolpert(1992)}]{wolpert1992stacked}
Wolpert, D.~H. 1992, Neural networks, 5, 241

\bibitem[{Wouthuysen(1952)}]{wouthuysen1952excitation}
Wouthuysen, S. 1952, The Astronomical Journal, 57, 31

\bibitem[{Xue {et~al.}(2022)Xue, Tong, \& Neri}]{xue2022ensemble}
Xue, Y., Tong, Y., \& Neri, F. 2022, Information Sciences, 608, 453

\bibitem[{Yaman {et~al.}(2018)Yaman, Subasi, \& Rattay}]{yaman2018comparison}
Yaman, M.~A., Subasi, A., \& Rattay, F. 2018, Symmetry, 10, 651

\bibitem[{Yamashita {et~al.}(2018)Yamashita, Nishio, Do, \& Togashi}]{yamashita2018convolutional}
Yamashita, R., Nishio, M., Do, R. K.~G., \& Togashi, K. 2018, Insights into imaging, 9, 611

\bibitem[{Yang {et~al.}(2015)Yang, Xie, Yuan, Zdziarski, Gierli{\'n}ski, Ho, \& Yu}]{yang2015correlation}
Yang, Q.-X., Xie, F.-G., Yuan, F., {et~al.} 2015, Monthly Notices of the Royal Astronomical Society, 447, 1692

\bibitem[{Yang {et~al.}(2020)Yang, Yu, \& Zhou}]{yang2020lstm}
Yang, S., Yu, X., \& Zhou, Y. 2020, in 2020 International workshop on electronic communication and artificial intelligence (IWECAI), IEEE, 98--101

\bibitem[{Yi{\u{g}}it \& Amasyali(2021)}]{yiugit2021simple}
Yi{\u{g}}it, G. \& Amasyali, M.~F. 2021, in 2021 international conference on INnovations in intelligent SysTems and applications (INISTA), IEEE, 1--6

\bibitem[{Ying(2019)}]{ying2019overview}
Ying, X. 2019, in Journal of physics: Conference series, Vol. 1168, IOP Publishing, 022022

\bibitem[{Yoshida {et~al.}(2003)Yoshida, Sokasian, Hernquist, \& Springel}]{yoshida2003early}
Yoshida, N., Sokasian, A., Hernquist, L., \& Springel, V. 2003, The Astrophysical Journal, 591, L1

\bibitem[{Yu \& Liu(2003)}]{yu2003feature}
Yu, L. \& Liu, H. 2003, in Proceedings of the 20th international conference on machine learning (ICML-03), 856--863

\bibitem[{Zahn {et~al.}(2007)Zahn, Lidz, McQuinn, Dutta, Hernquist, Zaldarriaga, \& Furlanetto}]{zahn2007simulations}
Zahn, O., Lidz, A., McQuinn, M., {et~al.} 2007, The Astrophysical Journal, 654, 12

\bibitem[{Zargar(2021)}]{zargar2021introduction}
Zargar, S. 2021, Department of Mechanical and Aerospace Engineering, North Carolina State University, Raleigh, North Carolina, 27606

\bibitem[{Zaroubi(2012)}]{zaroubi2012epoch}
Zaroubi, S. 2012, The first galaxies: theoretical predictions and observational clues, 45

\bibitem[{Zeng {et~al.}(2022)Zeng, Ma, Wang, \& Cui}]{zeng2022parking}
Zeng, C., Ma, C., Wang, K., \& Cui, Z. 2022, Ieee Access, 10, 47361

\bibitem[{Zygelman(2005)}]{zygelman2005hyperfine}
Zygelman, B. 2005, The Astrophysical Journal, 622, 1356

\end{thebibliography}

\end{document}